\documentclass[journal]{vgtc}                

\usepackage{microtype}                 
\PassOptionsToPackage{warn}{textcomp}  
\usepackage{textcomp}                  
\usepackage{times}                     
         
\usepackage{cite}                      
\usepackage{float}

\onlineid{0}

\vgtccategory{Research}

\vgtcpapertype{application/design study}

\title{\textsc{\textbf{MelodyVis}}: Visual Analytics for Melodic Patterns in Sheet Music}

\author{Matthias Miller, Daniel Fürst, Maximilian T. Fischer, Hanna Hauptmann, Daniel Keim, and Mennatallah El-Assady}

\authorfooter{
  \item \authororcid{Matthias Miller}{0000-0002-6281-2173} and \authororcid{Menna El-Assady}{0000-0001-8526-2613} are with \href{https://ivia.ch/}{ETH IVIA Lab}, Zurich. 
  \item \authororcid{Daniel Fürst}{0000-0002-0407-2867}, \authororcid{Maximilian T. Fischer}{0000-0001-8076-1376}, and \authororcid{Daniel A. Keim}{0000-0001-7966-9740} are with University of Konstanz. 
  \item \authororcid{Hanna Hauptmann}{0000-0002-6840-5341} is with Utrecht University, The Netherlands. 	
}

\usepackage{xspace}
\newcommand{\toolname}{\href{http://visual-musicology.com/melodyvis/}{\textsc{MelodyVis}}\xspace} 
 
\abstract{
Manual melody detection is a tedious task requiring high expertise level, while automatic detection is often not expressive or powerful enough. 
Thus, we present \toolname, a visual application designed in collaboration with musicology experts to explore melodic patterns in digital sheet music. 
\toolname features five connected views, including a Melody Operator Graph and a Voicing Timeline. The system utilizes eight atomic operators, such as transposition and mirroring, to capture melody repetitions and variations. 
Users can start their analysis by manually selecting patterns in the sheet view, and then identifying other patterns based on the selected samples through an interactive exploration process.
We conducted a user study to investigate the effectiveness and usefulness of our approach and its integrated melodic operators, including usability and mental load questions. 
We compared the analysis executed by 25 participants with and without the operators. 
The study results indicate that the participants could identify at least twice as many patterns with activated operators. 
\toolname allows analysts to steer the analysis process and interpret results.
Our study also confirms the usefulness of \toolname in supporting common analytical tasks in melodic analysis, with participants reporting improved pattern identification and interpretation. 
Thus, \toolname addresses the limitations of fully-automated approaches, enabling music analysts to step into the analysis process and uncover and understand intricate melodic patterns and transformations in sheet music.
}


\teaser{
  \vspace*{.5em}
  \centering
  \href{http://visual-musicology.com/melodyvis/}{\includegraphics[width=\linewidth]{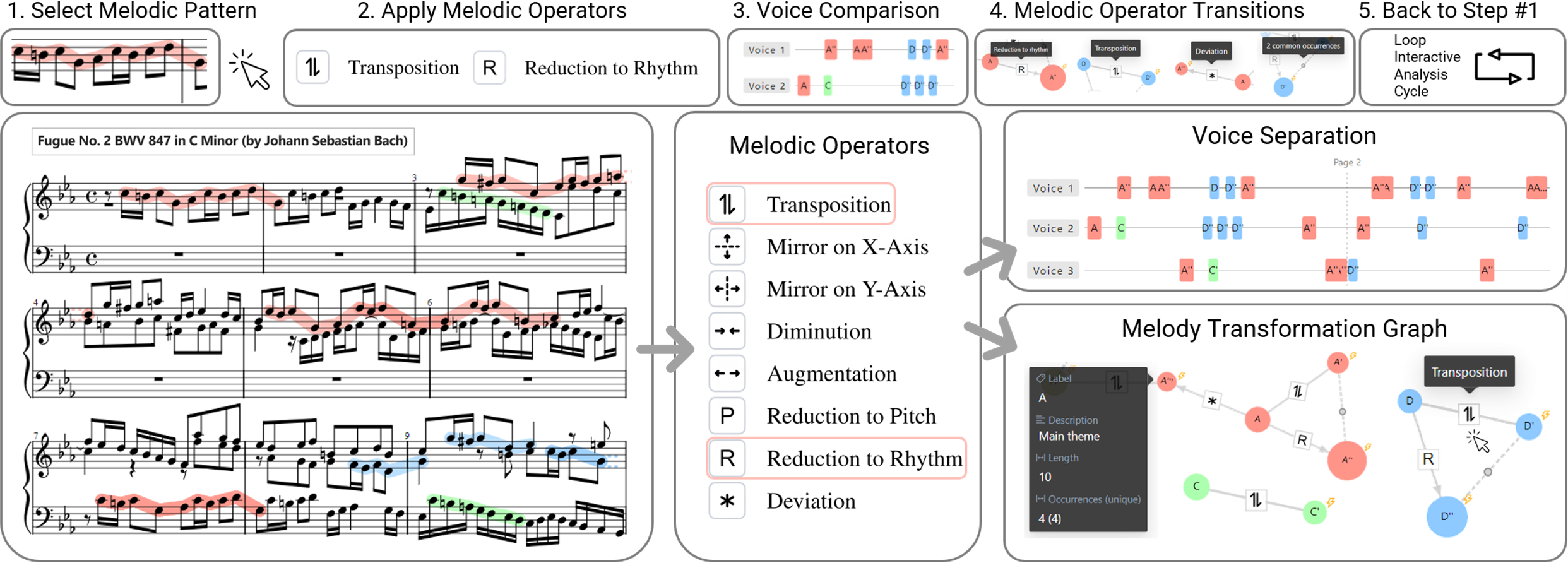}}    
  \caption{%
    \toolname provides an interactive workflow for the analysis of melodic patterns in sheet music. 
    Analysts can interactively select a pattern (1) that is used to apply melodic operators (2) leading to further highlights in the \emph{sheet view}.
    Then, the \emph{Voice Separation View} (3) provides an overview and distribution of the different voices, and the \emph{Melody Transformation Graph} (4) supports analyzing melodic variations.
    Finally, the analyst can continue adding further patterns by looping through the analysis process (5).     
  }
  \label{fig:teaser}
}

\graphicspath{{figs/}{figures/}{pictures/}{images/}{./}} 

\usepackage{tabu}                      
\usepackage{booktabs}                  
\usepackage{lipsum}                    
\usepackage{mwe}                       
\usepackage{mathptmx}                  
\usepackage{CJKutf8}

\usepackage[usenames,dvipsnames,table,svgnames]{xcolor}
\usepackage{tcolorbox}
\usepackage{svg}
\usepackage{adjustbox}
\usepackage{amssymb}
\usepackage{placeins}
\tcbuselibrary{skins}

\definecolor{pitch_c}{rgb}{0.976, 0.031, 0.349}
\definecolor{pitch_d}{rgb}{0.969, 0.576, 0.000}
\definecolor{pitch_e}{rgb}{0.969, 0.957, 0.035}
\definecolor{pitch_f}{rgb}{0.898, 0.031, 0.965}
\definecolor{pitch_g}{rgb}{0.973, 0.259, 0.012}
\definecolor{pitch_a}{rgb}{0.965, 0.816, 0.035}
\definecolor{pitch_b}{rgb}{0.835, 0.961, 0.000}
\definecolor{pitch_fis}{rgb}{0.247, 0.976, 0.008}
\definecolor{pitch_cis}{rgb}{0.024, 0.969, 0.659}
\definecolor{pitch_gis}{rgb}{0.024, 0.439, 0.969}

\usepackage{graphicx}
\usepackage{capt-of}
\usepackage{wasysym}

\usepackage{amssymb}

\usepackage{xspace}

\usepackage{enumitem}

\usepackage{tabu}

\usepackage{picinpar}

\usepackage[utf8]{inputenc}

\usepackage{microtype}
\usepackage{wrapfig}
\usepackage{enumitem}
\usepackage{soul}
\usepackage{graphbox}

\usepackage{calc}
\newlength\myheight
\newlength\mydepth
\settototalheight\myheight{Xygp}
\catcode`\@ = 11
\newdimen\@InsertBoxMargin
\newcount\@numlines    
\newcount\@linesleft   
\def\ParShape{%
    \@numlines = 0
    \def\@parshapedata{ }
    \afterassignment\@beginParShape
    \@linesleft
}%
\def\@beginParShape{%
    \ifnum \@linesleft = 0
      \let\@whatnext = \@endParShape
    \else
      \let\@whatnext = \@readnextline
    \fi
    \@whatnext
}%
\def\@endParShape{%
    \global\parshape = \@numlines \@parshapedata
}%
\def\@readnextline#1 #2 #3 {
    \ifnum #1 > 0
      \bgroup  
        \dimen0 = \hsize
        \advance \dimen0 by -#2  
        \advance \dimen0 by -#3  
        \count0 = 0
        \loop
          \global\edef\@parshapedata{%
            \@parshapedata    
            #2                
            \space            
            \the\dimen0       
            \space            
          }%
          \advance \count0 by 1
          \ifnum \count0 < #1
        \repeat
      \egroup
      \advance \@numlines by #1
    \fi
    \advance \@linesleft by -1
    \@beginParShape
}%
\newbox\@boxcontent     
\newcount\@numnormal    
\newdimen\@framewidth   
\newdimen\@wherebottom  
\newif\if@byframe       
\@byframefalse
\def\InsertBoxC#1{%
  \leavevmode
  \vadjust{
    \vskip \@InsertBoxMargin
    \hbox to \hsize{\hss#1\hss}
    \vskip \@InsertBoxMargin
  }%
}%
\def\InsertBoxL#1#2{%
  \@numnormal = #1
  \setbox\@boxcontent = \hbox{#2}%
  \let\@side = 0
  \futurelet \@optionalparameter \@InsertBox
}
\def\InsertBoxR#1#2{%
  \@numnormal = #1
  \setbox\@boxcontent = \hbox{#2}%
  \let\@side = 1
  \futurelet \@optionalparameter \@InsertBox
}%
\def\@InsertBox{%
  \ifx \@optionalparameter [
    \let\@whatnext = \@@InsertBoxCorrection
  \else
    \let\@whatnext = \@@InsertBoxNoCorrection
  \fi
  \@whatnext
}%
\def\@@InsertBoxCorrection[#1]{%
  \ifx \@side 0
    \@@InsertBox{#1}{0}{{\the\@framewidth} 0cm}%
  \else
    \@@InsertBox{#1}{1}{0cm {\the\@framewidth}}%
  \fi
}%
\def\@@InsertBoxNoCorrection{%
  \@@InsertBoxCorrection[0]%
}%
\def\@@InsertBox#1#2#3{%
  \MoveBelowBox
  \@byframetrue
  \@wherebottom = \baselineskip
  \multiply \@wherebottom by \@numnormal
  \advance \@wherebottom by 2\@InsertBoxMargin
  \advance \@wherebottom by \ht\@boxcontent
  \advance \@wherebottom by \pagetotal
  \ifdim \pagetotal = 0cm
    \advance \@wherebottom by -\baselineskip  
  \fi
  \advance \@wherebottom by #1\baselineskip
  \@framewidth = \wd\@boxcontent
  \advance \@framewidth by \@InsertBoxMargin
  \bgroup  
    \ifdim \pagetotal = 0cm
      \dimen0 = \vsize
    \else
      \dimen0 = \pagegoal
    \fi
    \ifdim \@wherebottom > \dimen0
      \immediate\write16{+--------------------------------------------------------------+}%
      \immediate\write16{| The box will not fit in the page. Please, re-edit your text. |}%
      \immediate\write16{+--------------------------------------------------------------+}%
      \vrule width \overfullrule
    \fi
  \egroup
  \prevgraf = 0
  \vbox to 0cm{%
    \dimen0 = \baselineskip
    \multiply \dimen0 by \@numnormal
    \advance \dimen0 by -\baselineskip
    \setbox0 = \hbox{y}%
    \vskip \dp0
    \vskip \dimen0
    \vskip \@InsertBoxMargin
    \ifnum #2 = 1
      \vtop{\noindent \hbox to \hsize{\hss \box\@boxcontent}}%
    \else
      \vtop{\noindent \box\@boxcontent}%
    \fi
    \vss
  }%
  \vglue -\parskip
  \vskip -\baselineskip
  \everypar = {%
    \ifdim \pagetotal < \@wherebottom
      \bgroup  
        \dimen0 = \@wherebottom
        \advance \dimen0 by -\pagetotal
        \divide \dimen0 by \baselineskip
        \count1 = \dimen0
        \advance \count1 by 1
        \advance \count1 by -\@numnormal
        \ifnum #2 = 1
          \ParShape = 3
                      {\the\@numnormal}   0cm   0cm
                      {\the\count1}       0cm   {\the\@framewidth}
                      1                   0cm   0cm
        \else
          \ParShape = 3
                      {\the\@numnormal}   0cm                  0cm
                      {\the\count1}       {\the\@framewidth}   0cm
                      1                   0cm                  0cm
        \fi
      \egroup
    \else
      \@restore@    
    \fi
  }%
  \def\par{%
      \endgraf
      \global\advance \@numnormal by -\prevgraf
      \ifnum \@numnormal < 0
        \global\@numnormal = 0
      \fi
      \prevgraf = 0
  }%
}%
\def\MoveBelowBox{%
  \par
  \if@byframe
    \global\advance \@wherebottom by -\pagetotal
    \ifdim \@wherebottom > 0cm
      \vskip \@wherebottom
    \fi
    \@restore@
  \fi
}%
\def\@restore@{%
    \global\@wherebottom = 0cm
    \global\@byframefalse
    \global\everypar = {}%
    \global\let \par = \endgraf
    \global\parshape = 1 0cm \hsize
}%
\ifx \documentclass \@Dont@Know@What@It@Is@
\else
  \let \pageno = \c@page
\fi

\catcode`\@ = 12

\usepackage{hyphenat}

\usepackage[english]{babel}
\addto\extrasenglish{%
}

\providecommand*\comment[1]{}
\let\commentstart=\iffalse

\providecommand*\thisisacomment[1]{}
\let\thisisacommentstart=\iffalse

\newcommand{\subhead}[1]{\vspace{2pt} \noindent \textbf{#1}}

\newcommand{\studyQuote}[1]{``\emph{#1}''}

\newcommand{\ignore}[1]{
}

\newcommand{\tss}{\textsuperscript}

\usepackage{tikz}
\tikzstyle{mybox}=[draw=blue, fill=orange!20, align=left,  text width = 0.45\linewidth]

\colorlet{shadecolor}{lightgray!40}
\newcommand{\figbordercolor}{shadecolor}
\newcommand{\figpaddingcolor}{white}
\newcommand{\addbordertofigure}[1]{
\fcolorbox{\figbordercolor}{\figpaddingcolor}{#1}
}

\newcommand{\wrapfigcustom}[7]{
\begin{wrapfigure}[#1]{#2}{#3}
 \vspace{#6}
  \begin{center}
    \hspace*{#7}\fcolorbox{\figbordercolor}{\figpaddingcolor}{\includegraphics[width=#4]{#5}}
  \end{center}
\end{wrapfigure}
}

\begin{document}

\firstsection{Introduction}

\maketitle

For a very long time, music has been an expression of personality, cultural identity, and a universal factor in religious traditions~\cite{hudson2006regions,hudson2017church}.
Thus, it has been integral to society, human culture, and individual experiences~\cite{cross2001music}.
Especially the melodic character plays a pervasive role, possessing the remarkable ability to stir emotion, create social bonds, and enhance cognitive processes when people engage with music~\cite{weisgerber2013music}.
Throughout time, different musical trends developed across many cultures~\cite{savage2020toward}.
The importance of music and its overall cultural evolution for humans is estimated to have started around 43 to 50 thousand years ago~\cite{merker1999synchronous}.
For a long time, music was only preserved through oral traditions~\cite{hood1959reliability}.
Apart from physical artifacts and depictions of musicians and instruments, the oldest known musical notation, dating back to approximately 1,400 BCE, was discovered on a Sumerian clay tablet in modern-day Syria, near Ras al-Shamra, the ancient city of Ugarit~\cite{guterbock1970musical,dumbrill2005archaeomusicology}. 
The tablet features engraved melodies, including Hurrian Hymn No. 6, also known as the ``Seikilos Epitaph'' which historians consider the earliest complete surviving musical composition in a notated form.
Music notations have been developed to pass musical knowledge on to future generations and composed hymns and melodies as part of the cultural identity.
Well-known is the wenzi pu %
\begin{CJK*}{UTF8}{bsmi}
(文字譜)
\end{CJK*} %
notation for the guqin instrument, common in seventh-century China.
More common to many today is the Western Common Music Notation~\cite{Schottstaedt1997CMN}. 

The advantage of visual representation of music is the extraction of detailed musical notation from auditorial information that is subject to interpretation by a singer or musician~\cite{cardew_1961_notationinterpretation}.
Musical notation, in particular sheet music~\cite{elliker1999toward}, provides a precise representation of a composition, allowing for a more accurate investigation of patterns and themes compared to audio recordings~\cite{lassfolk2004music}.
Both formats describe musical information on different semantic levels. 
Sheet music comprises abstract parameters such as notes, keys, measures, and other instructions in a visual form, providing a detailed visual explanation of how a musical composition should be performed.
In contrast, audio recordings result from music in an acoustic form comprising detailed nuances in interpretation and expression, which are generally not written down but make music ``come alive''~\cite{fremerey2009sheetaudiomusic}.
Thus, analyzing patterns in sheet music facilitates the study of a composer’s intentions and the structural elements inherent in a piece, while audio analysis involves the interpretation of a musician, which may introduce deviations or inaccuracies~\cite{fremerey2009sheetaudiomusic}.
Moreover, sheet music captures essential musical elements, such as pitch, rhythm, dynamics, and articulation~\cite{miller_analyzing_2018}, in a visually accessible format that simplifies pattern identification and comparison across different compositions from a larger corpus~\cite{corpusvis2022miller}. 
The notational unambiguity of sheet music enables researchers to examine subtle variations in melodic and harmonic structures and uncover relationships between different pieces or composers that may not be readily apparent in audio recordings.
For this reason, most musicologists prefer written (sheet) music to analyze the meaning of music due to its compactness and explicitness~\cite{dudeque2005music}.

Melodic themes, characterized by reoccurring note sequence patterns and their variations in sheet music, are often a composition's most salient and memorable aspects and of specific interest to musicologists~\cite{shih2001dictionary}. 
Analyzing sheet music for melodic patterns is also crucial for assessing the originality of melodic content~\cite{mullensiefen2009court}, understanding stylistic and genre-specific attributes~\cite{dannenberg2010style}, and gaining insights into historical and cultural contexts.
Integrating the analyst into the analysis process through Visual Analytics (VA) is crucial because it capitalizes on human expertise and intuition~\cite{framingvisualmusicology2019}, addressing the limitations of fully automated methods that may overlook subtle, context-specific or complex patterns in the music~\cite{wolkowicz2013application}.
Visual analysis of sheet music allows for examining multiple voices simultaneously, enabling a comprehensive understanding of the interplay between various elements within a composition. 
In contrast, the analysis of audio recordings may struggle to disentangle overlapping voices~\cite{chew2004separating}, making it difficult to discern individual patterns and melodic relationships.
Focusing on visual analysis of sheet music data improves the preservation of a composer’s intent, enhancing pattern identification accuracy and facilitating the examination of complex musical interrelationships. 
This approach makes sheet music more suitable for in-depth melodic analysis, thus facilitating analysts' understanding of musical compositions~\cite{MillerFHKE22Augmenting}.

In music analysis, melodic transformations are crucial in understanding how melodic material is processed and developed within a composition~\cite{dowling1972recognition}. 
Researchers can gain insights into composers' employed techniques by focusing on such transformations.
Studying these transformations on sheet music instead of audio recordings ensures a more precise and accurate examination of the underlying melodic structures and relationships. 
Manually identifying these patterns and their transformations in sheet music can be challenging and require music analysts to understand music notation thoroughly~\cite{miller_analyzing_2018}. 
Visual Analytics offers a promising solution for visually analyzing melodic patterns in sheet music. 
By integrating the analyst into the process, VA facilitates a more transparent and understandable analysis process compared to fully automated methods~\cite{framingvisualmusicology2019}.
It can support learning and interpretation tasks and enable music analysts to explore and uncover intricate patterns and relationships that might otherwise remain hidden~\cite{MillerFHKE22Augmenting}.

To address the issue of how information visualization research can contribute to melody analysis, we consider the research question: \emph{How can we employ Visual Analytics to enable the semi-automatic investigation of melodic patterns based on sheet music?}
By focusing on melodic operators, which are transformations that manipulate the melodic material of music compositions, we designed \toolname, an interactive analysis interface that facilitates the visual investigation of melodic content throughout a musical composition.

\subhead{Contributions --} 
This work contributes \toolname, a visual interactive application that encompasses multiple seamlessly connected components based on an interactive workflow for analyzing melodic patterns. 
We offer a comprehensive problem characterization for melodic pattern analysis, addressing visualization requirements in the context of sheet music data. 
We also provide an overview of potential melody analysis tasks and describe typical users for sheet music melody analysis. 
We conducted both qualitative and quantitative evaluations to assess the applicability and effectiveness of \toolname, discussing its current potential and remaining limitations. 
Finally, we present potential future research directions to inspire interdisciplinary collaboration between the information visualization community and the field of musicology, with a special focus on melody analysis.

\section{Background and Related Work} 

Music analysts investigate and explore melodic themes in sheet music to understand musical compositions~\cite{chintaka2004content}.
Prior works by Miller et al.~\cite{miller_augmenting_2019,MillerFHKE22Augmenting} and Fürst et al.~\cite{fuerst_augmenting_2020} demonstrated the value of augmenting digital sheet music to enhance the analysis of the structure and progression of harmony and rhythm. 
On a more abstract level, Sapp used tonal material of compositions to enable visual analysis of key modulation providing KeyScapes~\cite{sapp_visual_2005}. 
The introduction of numerous digital formats for musical scores, including MIDI~\cite{helio2017understanding}, MusicXML~\cite{good_musicxml_2001}, Humdrum \texttt{**kern}~\cite{huron2002musicinformation}, Lilypond~\cite{DBLP:conf/ismir/SinclairDF06}, ABC notation~\cite{abcnotation1991walshaw}, MEI~\cite{roland2002music}, and GUIDO notation~\cite{Hoos1998TheGN}, and the technological progress enabled musicologists to take advantage of computational approaches that use algorithms to extract valuable information from musical pieces automatically~\cite{hartmann2007interactive}.

In this work, we specifically focused on investigating melodic sequences based on digital sheet music by employing Visual Analytics.
Subsequently, we discuss the primary visual approaches that facilitate melodic analysis, focusing on the existing computational methods. 
Then, we provide an overview of published applications that assist in melodic information retrieval and the analysis of recurring patterns.

\subsection{Sheet Music Retrieval based on Melodic Sequences}

\ignore{
\begin{itemize}
    \item Global Chant~\cite{kolycek2009globalchant}: enter at least two notes - no operators 
    \item \href{https://hymnary.org/melody/search}{Hymnary: Search by given melody (transposition is applied}
    \item Burghardt and Lamm~\cite{burghardt2017entwicklung}
    \item Fast Melody Finder~\cite{rho2004fmf}
\end{itemize}
}

In the related work, various approaches aim to retrieve musical compositions from a larger dataset based on melodic content, employing diverse methods to search for sheet music. 
These methods can be categorized into pattern-matching notes, filtering by genre or duration, n-gram approaches, contour-based searches, and multimedia queries.

For example, Koláček introduced a simple note query based on the CWMN, where a user can easily add a few notes to search the \emph{Global Chant Database}~\cite{kolycek2009globalchant}. 
He does not provide more sophisticated melodic transformations to identify variations. 
Instead, a few additional fields allow for query text or genre information. 
The melody query input provided on \href{https://hymnary.org/}{Hymnary.org} is similar~\cite{schneider2012hymnary}. 
Still, it has a few additional options to define the duration of notes and also an interactive keyboard to facilitate selecting the pitch for a query.
Dig That Lick provides a similar pattern similarity search interface for different Jazz music databases introduced by Henry et al.~\cite{henry2021dig}.
Burghardt and Lamm use n-gram matching with a specific note and a relative interval search to identify motif transpositions. 
Still, they do not enable further operators or the direct selection of melodic examples from a piece~\cite{burghardt2017entwicklung}.
They also provide an option similar to Musipedia~\cite{irwin2008musipedia} where an analyst can use the Parsons Code~\cite{parsons1975dictionary}, also known as pitch contour~\cite{melody2012salamon}.
Themefinder by Kornstädt separates the contour query into a gross and refined contour query to enable slight melodic nuances in the query in addition to an absolute pitch and relative interval query~\cite{kornstadt1998themefinder}.
Musipedia provides some flexibility with a keyboard search, rhythm search, and microphone/hum query to look for music pieces. \\
Analyzing the relationship of musical patterns inside a single composition is not possible, while a user can not select a query template based on a given composition~\cite{irwin2008musipedia}.
Fast Melody Finder by Rho and Hwang synthesizes an audio input to create a pitch contour query string to search a database for matches addressing singing and humming search tasks~\cite{rho2004fmf}. 
A significant drawback of most approaches is that the user does not know whether there are any results before executing the query, leading to a tedious trial-and-error approach.

\subsection{Melodic Pattern Mining and Extraction}
Besides music information retrieval tasks, some approaches address melodic pattern mining based on a given data set.
In this scenario, the users don't need to provide a specific search template in advance. 
Still, rather algorithmic approaches identify salient and frequent melodic patterns potentially interesting to a music analyst.
For example, Frieler et al. used Melodia~\cite{melody2012salamon}, a state-of-the-art melody estimation method to extract jazz solo melodies from the audio-based Weimar Jazz Database~\cite{frieler2019don}.
Wen et al. created a corpus of compositions from Haydn, Mozart, and Beethoven using the MusicXML format~\cite{good_musicxml_2001} to perform melodic feature extraction and then applying a k-means clustering on top of it~\cite{wen2017musicstyle}.
To identify the similarity of the main melodies of cover songs of popular songs, Tsai et al.\ introduced a method to match the main theme based on audio data~\cite{tsai2008usingsimilartymainmelody}.\\
Rigaux and Travers introduced and designed a scalable search engine for melodic queries based on \textsc{Elastic-Search} for collection indexing employing variant matching~\cite{rigaux2019scalable}.
Their work disembogues in the approach by Zhu et al. in an extensive framework for feature extraction and a text-oriented search engine~\cite{zhu2022framework} to manage extensive collections of digitized music documents.
A drawback of these approaches is the inability of a user to directly explore and steer the results, as they only perform an algorithmic matching of which the results are not further analyzable by a musicologist on a lower level.
They don't leverage a visual interactive interface to make the search results analyzable within the context of its finding.
Rather, they focus on their approach's overall accuracy in terms of statistically measuring the quality of the identified results compared to a ground truth data set.
Soum-Fontez et al. investigated how textural characteristics can help identify melodic voices in polyphonic music, explicitly focusing on string quartets~\cite{soum2021symbolic}.
Our approach follows a similar approach to separate multiple voices into disjoint voices based on envelope extraction~\cite{gray2020fromnotelevel}.

Various approaches exist for melody analysis in sheet music.
Meek et al. provide an automatic thematic extractor to identify \textit{themes} by automatically extracting melodic patterns while reducing the resulting space through suitable filters~\cite{meek03automthematicextractor}.
Urbano proposes the library \studyQuote{MelodyShape} that enables similarity search for melodies based on the shape~\cite{Urbano2015MelodyShapeA} but not on the sheet level. 
Unfortunately, these approaches reside either on the conceptual level or require programming skills that musicologists, domain experts, music teachers, and students might not have. 
Thus, these approaches are not directly accessible to a broad audience that requires intuitive and interactive visualization. 
Miller et al. presented a first visual approach to enable direct melody analysis for sheet music data, but their melody query only allows for exact matching, transpositions, and contour search~\cite{MillerFHKE22Augmenting}.
While we consider this a first step in the right direction, we identified that none of the existing visual approaches allows for a voice-separated investigation of melodic transformations.
Only the Verovio Humdrum Viewer by Sapp already provides a set of melodic analysis functions including imitation, melisma, dissonance labels, composite rhythm, and meter analysis with limited opportunity for visual investigation~\cite{sapp2017verovio}.
Therefore, we aim to address this research gap by leveraging information visualization techniques to support the melodic analysis using multiple components that support low and high-level analysis tasks.

\subsection{Applying Methodology Transfer}

This work is inspired by concepts from the text analytics domain within the research field of digital humanities. 
\toolname is a visualization system for exploring the relationships of melody patterns in a musical composition in sheet music, inspired by the methodology transfer from Poemage~\cite{poemage2016curdy}, a system designed for analyzing the sonic patterns of poems.
This is similar to the methodology transfer use case suggested by Miller et al.~\cite{framingvisualmusicology2019}.
The need for close reading~\cite{janicke_close_2015} in melody analysis is essential, as it enables the in-depth examination of melodic patterns and relationships within sheet music, closely aligning with the manual analysis process musicologists traditionally employ. 
As we already discussed, melody analysis examines the complex structures formed by the interaction of melodic patterns — musical phrases connected through melodic transformations, such as transpositions or inversions — throughout the composition. 
\toolname was developed as part of an ongoing, highly exploratory collaboration between data visualization experts and musicologists. 
This methodology transfer adapts the exploration of sonic patterns in textual data to the analysis of melodic patterns in sheet music, enabling users to efficiently explore, interpret, and understand complex melodic relationships while maintaining their domain knowledge and intuition in the analysis process.

Miller et al. introduced \href{https://visual-musicology.com/graph/}{a structured graph} within the field of \href{https://visual-musicology.com/}{Visual Musicology} that facilitates the classification of \toolname~\cite{framingvisualmusicology2019}.
In this work, we use the \emph{Structural Features } -- \emph{Pitch} and \emph{Rhythm} -- from the \emph{Data} branch and address the \emph{Analytical Tasks} -- \emph{Information Retrieval} and \emph{Contextual Analysis} within the \emph{domains} of \emph{Theory \& Analysis} and \emph{Education}.
\toolname considers the three \emph{Visualization Tasks}: \emph{Overview/Summarization}, \emph{Navigation/Exploration}, and \emph{Comparison}.

\section{Structure of Melodic Patterns}
\label{structuremelodicpatterns}

To analyze melodic patterns, it is essential to understand the meaning and relationship of \toolname's main building blocks:

\subhead{Melodic-related Musical Features --} 
Melody is a compound musical feature intricately woven from pitch and time dimensions~\cite{MillerFHKE22Augmenting}. 
Pitch refers to the perceived frequency of musical notes, and is encoded vertically on the sheet through high or low positions~\cite{weihs_music_2017}.
This element dictates the progression and contour of the notes in a melody. 
Time, the horizontal dimension, involves rhythm, which combines patterns of onsets and duration in a sequence of musical notes~\cite{fuerst_augmenting_2020}, contributing to the overall flow and progression of the melody. 
A note's length depends on the tempo and meter given for a piece. 
The note's \textit{onset} defines its temporal placement in a composition~\cite{toussaint2019geometry}. 
In contrast to notes, \textit{rests} are the explicit information of the absence of sound providing temporal structure information. 
By focusing on the interplay of pitch and rhythm, we can analyze a melodic line's characteristics, capturing the essence of this compound feature. 
While other elements such as harmony~\cite{miller_augmenting_2019}, articulation, and timbre~\cite{miller_analyzing_2018} are also related to melody, they will not be the primary focus of our analysis.

We clearly distinguish between a single melody and a chord progression. 
By our definition, any melodic instrument, such as the recorder, can perform a melody, which would not be possible with notes that must be played simultaneously. 
Further, we can define \textit{motifs} and their \textit{variations}.
Melody is the most salient feature as it is usually remembered by the listener, e.g., influencing catchy tunes. 
Especially the aspect of the leitmotif~\cite{bribitzer-stull_2015} is of high interest to music analysts, which is the leading motive that influences the central theme of a musical piece. 

\subhead{Voicing and Melodic Texture --} Combining multiple simultaneously played voices leads to diverse musical textures~\cite{huron1989characterizing}. 
We consider three melodic texture types: monophony, polyphony, and homophony~\cite{cambouropoulos2006voice}. \\
\emph{Monophony}, also known as monody, describes the succession of single notes of only one voice comprising a melodic line without accompanying voices~\cite{weihs_music_2017}. 
It is the most basic form of musical texture with little information in the vertical component often occurring in folk and early music~\cite{richter2015monophony}.
Melody analysis based on monophonic music focuses solely on the primary line, which forms the entire musical structure.\\
\emph{Polyphony} comprises multiple independent melodic lines or voices played simultaneously, creating a complex and layered texture as the different melodies intersect but still fit well together~\cite{weihs_music_2017}. 
Polyphony is a common feature of music from the Baroque and Classical periods and is also found in many other musical traditions worldwide.
Melody analysis for polyphonic music is more complex, as it entails understanding the interplay between these lines and their relationships in its context. \\
\emph{Homophony} features a main melody supported by harmonic accompaniment, emphasizing the relationship between melody and harmony. 
Homophony also describes music in which two or more voices synchronously move together in harmony~\cite{weihs_music_2017}.
Studying homophonic compositions requires analyzing both aspects.
Homophony is a prevalent texture in Western classical music and popular music.

In this work, we focus on monophony and polyphony by considering independent voices~\cite{chew2004separating} without addressing harmonic progressions.

\subhead{Melodic Transformation --}
Besides voicing and musical texture, melodic transformations are relevant in shaping the structure of melodic patterns and their development throughout a musical composition. 
These techniques enable composers to craft interesting connections between different themes, leading to a broad and dynamic melodic landscape. 
In the context of information visualization, the study of melodic transformations offers a unique opportunity to explore the underlying patterns within musical compositions, providing insights into the creative processes of composers to develop melodic material.

By focusing on the atomic melodic operators that form the basis of these transformations, researchers in the information visualization community can contribute to a deeper understanding of the various ways in which composers have utilized and combined these building blocks. 
Furthermore, this interdisciplinary collaboration can lead to the development of innovative visualization techniques and tools that reveal hidden relationships, commonalities, and differences between existing melodic themes, ultimately enhancing our ability to analyze, interpret, and appreciate the complexity of musical compositions.

\clearpage

\subsection{Atomic Melodic Operators}
\label{subsec:melodicoperators}

Melodic transformations in musical compositions are characteristic means composers and artists use to manipulate melodic material.
Different transformations can be applied to a musical motif to identify repetitive melodic themes and variations in musical compositions to process the original melodic material. 
In the following, we describe eight operators we extracted from the literature that musical composers commonly use:

\newcommand{\opCardTitle}{}
\newcommand{\opCardDescription}{}
\newcommand{\opCardIcon}{}
\newcommand{\opCardScore}{}
\newcommand{\opCardExample}{}

\newcommand{\operatorAbbrev}{OP}

\newcommand{\opA}{\texttt{[\operatorAbbrev1]}}
\newcommand{\opB}{\texttt{[\operatorAbbrev2]}}
\newcommand{\opC}{\texttt{[\operatorAbbrev3]}}
\newcommand{\opD}{\texttt{[\operatorAbbrev4]}}
\newcommand{\opE}{\texttt{[\operatorAbbrev5]}}
\newcommand{\opF}{\texttt{[\operatorAbbrev6]}}
\newcommand{\opG}{\texttt{[\operatorAbbrev7]}}
\newcommand{\opH}{\texttt{[\operatorAbbrev8]}}

\newcommand{\operatorCard}[6]{
    \begin{tcolorbox}[
        enhanced,
        colback=white,
        colframe=gray!60,
        boxrule=1pt,
        rounded corners,
        title={#1~#6},
        coltitle=black,
        colbacktitle=gray!20,
        halign title=center,
        fonttitle=\bfseries
    ]
        \vspace*{-3pt}
        \begin{tabular}{@{}c@{\hspace{10pt}}p{0.85\linewidth}@{}}
            \raisebox{-0.4\totalheight}[0pt][0pt]{
                \includegraphics[width=0.07\linewidth]{#2}    
            } &
            {\centering #3}
        \end{tabular}
        
        \vspace*{5pt}    
        \includegraphics[width=1\linewidth]{#4}    

        {\small #5}
    \end{tcolorbox}
}

\renewcommand{\opCardTitle}{Transposition}
\renewcommand{\opCardDescription}{Transposes pattern by fixed but arbitrary interval}
\renewcommand{\opCardIcon}{figs/operator_icons/transposition}
\renewcommand{\opCardScore}{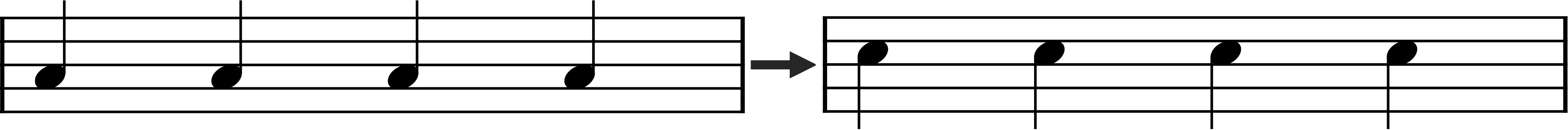}
\renewcommand{\opCardExample}{The transposition operator detects all patterns in which the pattern notes are shifted by a fixed interval while maintaining their relative positions.}
\operatorCard{\opCardTitle}{\opCardIcon}{\opCardDescription}{\opCardScore}{\opCardExample}{\opA}

We consider transposing melodic patterns as the most basic operator, as it is typically easy to identify such changes by simply listening to a song. 
Even detecting such occurrences of a given melodic theme in sheet music is the least difficult but already requires specific musical expertise, especially for polyphonic music. 

\renewcommand{\opCardTitle}{Mirror on X-Axis}
\renewcommand{\opCardDescription}{Mirrors pitches on x-axis through first pitch}
\renewcommand{\opCardIcon}{figs/operator_icons/mirror-x}
\renewcommand{\opCardScore}{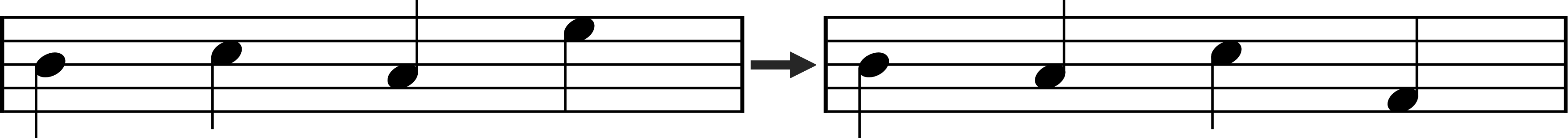}
\renewcommand{\opCardExample}{The x-axis-mirror operator identifies all patterns where the pitches of notes are reflected across the x-axis, using the first note’s pitch as a reference point.}
\operatorCard{\opCardTitle}{\opCardIcon}{\opCardDescription}{\opCardScore}{\opCardExample}{\opB}

Mirroring a melodic pattern based on the x-axis is also known as \emph{Inversion}~\cite{welker1982abstraction}. 
Composers often exploit this operator to create melodic variety by maintaining coherence and a sense of familiarity with the already-used theme material but presenting it in a new form. 
This technique is also common to establish unity within a composition but allows for further opportunities for exploring and developing a motif.

\renewcommand{\opCardTitle}{Mirror on Y-Axis}
\renewcommand{\opCardDescription}{Mirrors notes on y-axis of middle note(s)}
\renewcommand{\opCardIcon}{figs/operator_icons/mirror-y}
\renewcommand{\opCardScore}{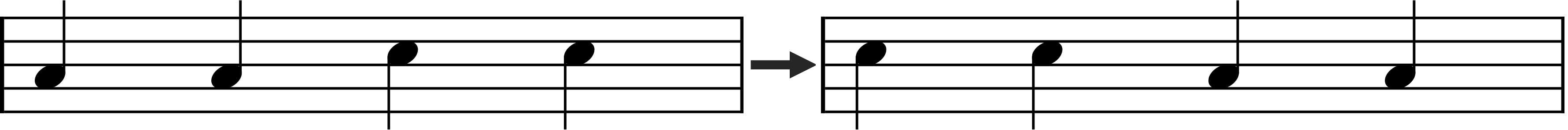}
\renewcommand{\opCardExample}{The mirror on the y-axis operator finds all patterns where the notes are reflected across the y-axis, using the middle note(s) of the pattern as a reference point.}
\operatorCard{\opCardTitle}{\opCardIcon}{\opCardDescription}{\opCardScore}{\opCardExample}{\opC}

The Y-Axis-Mirror operator is similar to \emph{Inversion}, and musicologists also call the operator \emph{Retrograding}~\cite{dowling1972recognition}.
This operator also leads to melodic variety and can introduce novelty, but it maintains a connection to the initial theme. 
Moreover, the two operators can be intentionally combined two create an even more intricate melodic landscape. 
For example, Arnold Schönberg extensively employed \emph{Inversion} and \emph{Retrograde} techniques in his twelve-tone compositions~\cite{carpenter2005piano}.

\renewcommand{\opCardTitle}{Diminution}
\renewcommand{\opCardDescription}{Shortens consistently the notes' length of a melody}
\renewcommand{\opCardIcon}{figs/operator_icons/diminution}
\renewcommand{\opCardScore}{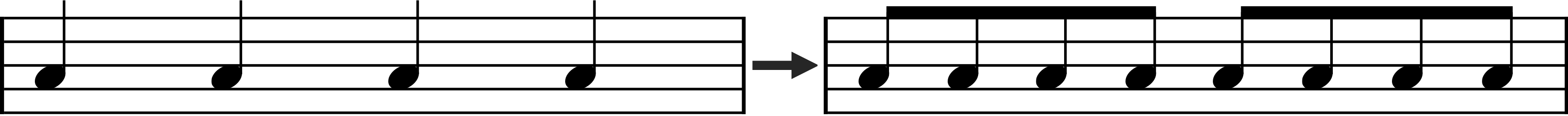}
\renewcommand{\opCardExample}{The diminution operator locates all patterns where the duration of notes is reduced by half, resulting in faster rhythmic patterns.}
\operatorCard{\opCardTitle}{\opCardIcon}{\opCardDescription}{\opCardScore}{\opCardExample}{\opD}

Diminuting melodic material also supports intensified excitement that can lead up to a climax in a composition~\cite{park2013piano}. 
While in the given example above, all the notes have the same duration, one can easily think of a less uniform rhythm of a melody. 
Then, applying \emph{Diminution} can create more sophisticated melodic lines through the proportional shortening of the note sequence.
While this technique has already been used in the Renaissance period, it continues to be a valuable composing tool for maintaining a coherent melodic core structure but introducing a sense of energy and momentum that can enhance the emotional impact or create a certain level of urgency. 
The operator is also used to develop contrapuntal textures, which add depth and complexity to compositions while showcasing a composer's creativity and overall compositional skill as has been employed by Händel~\cite{mann1962handels}, Bach~\cite{charlston2020patterns}, and Wagner~\cite{gilbert1926humor}.

\renewcommand{\opCardTitle}{Augmentation}
\renewcommand{\opCardDescription}{Proportionally elongates all notes of a melodic motif}
\renewcommand{\opCardIcon}{figs/operator_icons/augmentation}
\renewcommand{\opCardScore}{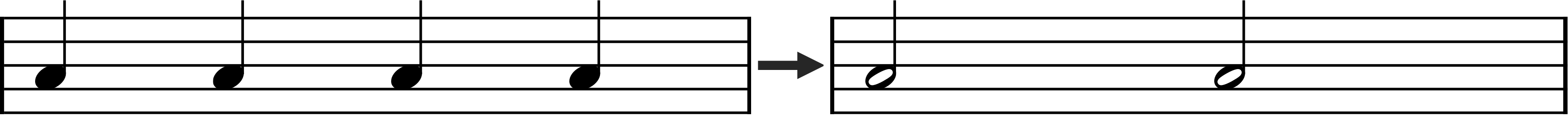}
\renewcommand{\opCardExample}{The augmentation operator uncovers all patterns where the duration of notes is increased by twice, resulting in slower rhythmic patterns.}
\operatorCard{\opCardTitle}{\opCardIcon}{\opCardDescription}{\opCardScore}{\opCardExample}{\opE}

As the opposite of \emph{Diminution}, \emph{Augmentation} leads to a proportional elongation of melodic lines.
Musicologists consider that diminution and augmentation enhance the overall structure and musical narrative by generating contrast between different composition sections. 
This operator is often applied to create accompanying melodic textures~\cite{learning2021wei}.

\renewcommand{\opCardTitle}{Reduction to Pitch}
\renewcommand{\opCardDescription}{Only considers pitches of pattern}
\renewcommand{\opCardIcon}{figs/operator_icons/reduction-to-pitch}
\renewcommand{\opCardScore}{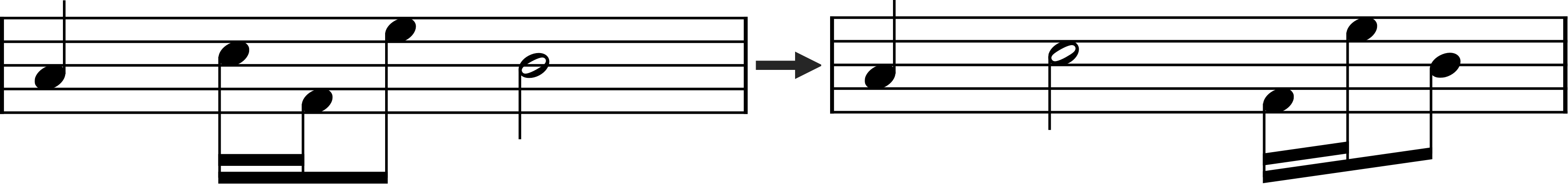}
\renewcommand{\opCardExample}{The reduction to pitch operator focuses on the pitch values of notes in the pattern, leading to various possible patterns based on pitch alone.}
\operatorCard{\opCardTitle}{\opCardIcon}{\opCardDescription}{\opCardScore}{\opCardExample}{\opF}

Only considering the pitch content facilitates focusing on the essential melodic structure, enabling the creation of less restricted and more digestible versions of an original theme. 
This operator also helps to clarify the core ideas of a more complex melody.
This technique allows for developing thematic material by applying the rhythmic structure of another melody, thus combining two previously unconnected motifs.

\renewcommand{\opCardTitle}{Reduction to Rhythm}
\renewcommand{\opCardDescription}{Only considers rhythm of pattern}
\renewcommand{\opCardIcon}{figs/operator_icons/reduction-to-rhythm}
\renewcommand{\opCardScore}{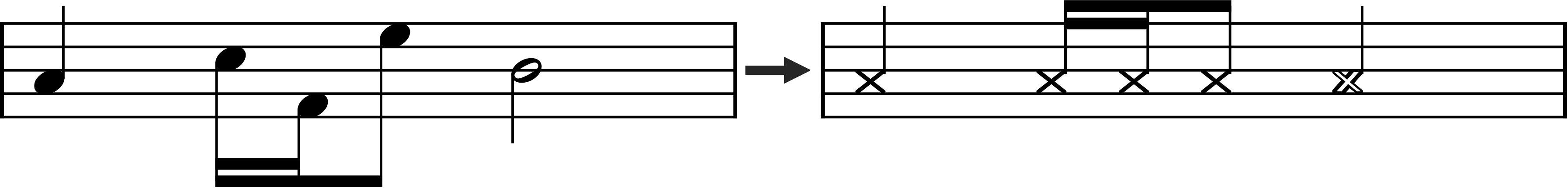}
\renewcommand{\opCardExample}{The reduction to rhythm operator emphasizes the duration values of notes in the pattern, leading to various possible patterns based on rhythm alone.}
\operatorCard{\opCardTitle}{\opCardIcon}{\opCardDescription}{\opCardScore}{\opCardExample}{\opG}

Similarly, only the rhythmic structure and content of an existing melody can be used to connect two unrelated melodic fragments. 
It can be used in a larger context, such as an opera, to connect two separate acting roles in a musical fashion by using their specific pitch material with the other rhythmic content, opening new ways of interpretation.

\renewcommand{\opCardTitle}{Deviation}
\renewcommand{\opCardDescription}{Permits up to 20 \% of all notes to deviate from pattern}
\renewcommand{\opCardIcon}{figs/operator_icons/deviation}
\renewcommand{\opCardScore}{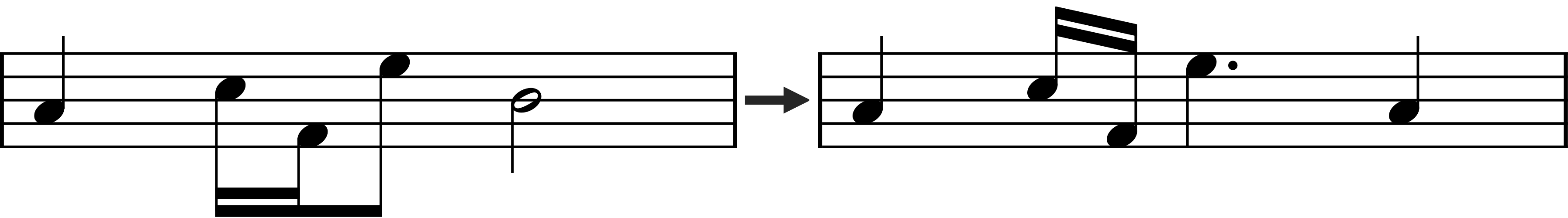}
\renewcommand{\opCardExample}{ Deviation allows a fixed percentage of notes (e.g., 20\%) in the pattern to differ, leading to various possible patterns based on the degree of deviation.}
\operatorCard{\opCardTitle}{\opCardIcon}{\opCardDescription}{\opCardScore}{\opCardExample}{\opH}

This operator considers deviation of rhythm and pitch allowing music analysts to identify how far a composer departed from melodic themes.
The \emph{Deviation} operator is especially useful to be more flexible in a search for similarities without the need to specify exact parameters.

\vspace{1cm}

\subsection{Complex Chaining of Melodic Operators}
\label{subsec:operatorchaining}
The atomic melodic operators described in Subsection~\ref{subsec:melodicoperators} provide a foundation for identifying and understanding individual melodic transformations in musical compositions. 
In practice, composers often apply multiple transformations simultaneously or consecutively, creating complex melodic variations. 
To capture such relationships and transformations, we can chain and combine the atomic operators to create complex melodic operators.
Chaining operators involve applying one operator after another, where the order of the operators influences the types of patterns identified. 
For instance, applying Transposition~\opA followed by Mirror on X-Axis~\opB would result in a different pattern than applying them in the reverse order. 
By chaining operators, we can uncover more subtle and sophisticated melodic variations, gaining a deeper understanding of a composer's creative process and the underlying structure of the composition.
Similarly, combining operators involves applying multiple operators simultaneously. 
This process can reveal patterns that exhibit multiple transformations at once, such as an Inversion combined with a Retrograde. 
In some cases, applying the same operator multiple times could serve as an identity function, which might not necessarily result in meaningful combinations. 
However, this flexibility allows for the discovery of a wider range of patterns, offering a more comprehensive analysis of melodic transformations.

There are benefits of complex operator chaining. 
(1) Increased flexibility: Complex operators allow analysts to discover a broader range of melodic patterns and transformations, resulting in a more complete understanding of a composition's structure and development.
(2) Enhanced understanding of compositional techniques: By uncovering subtle melodic variations, analysts can gain insights into a composer's creative process and intentions.
Likewise, there are also potential drawbacks to consider.
(1) Increased complexity: As the number of chained or combined operators increases, the complexity of the analysis grows, which could make it more challenging for analysts to interpret the results.
(2) Risk of overfitting: When searching for patterns using complex operators, there's a risk of identifying patterns that are not musically meaningful, especially if the analyst broadens the operators' parameter space too much.

\subsection{Analysis Tasks and Requirements}
\label{subsection:analysis_tasks}

\begin{figure}[h]
    \centering
    \addbordertofigure{
    \includegraphics[width=0.96\linewidth]{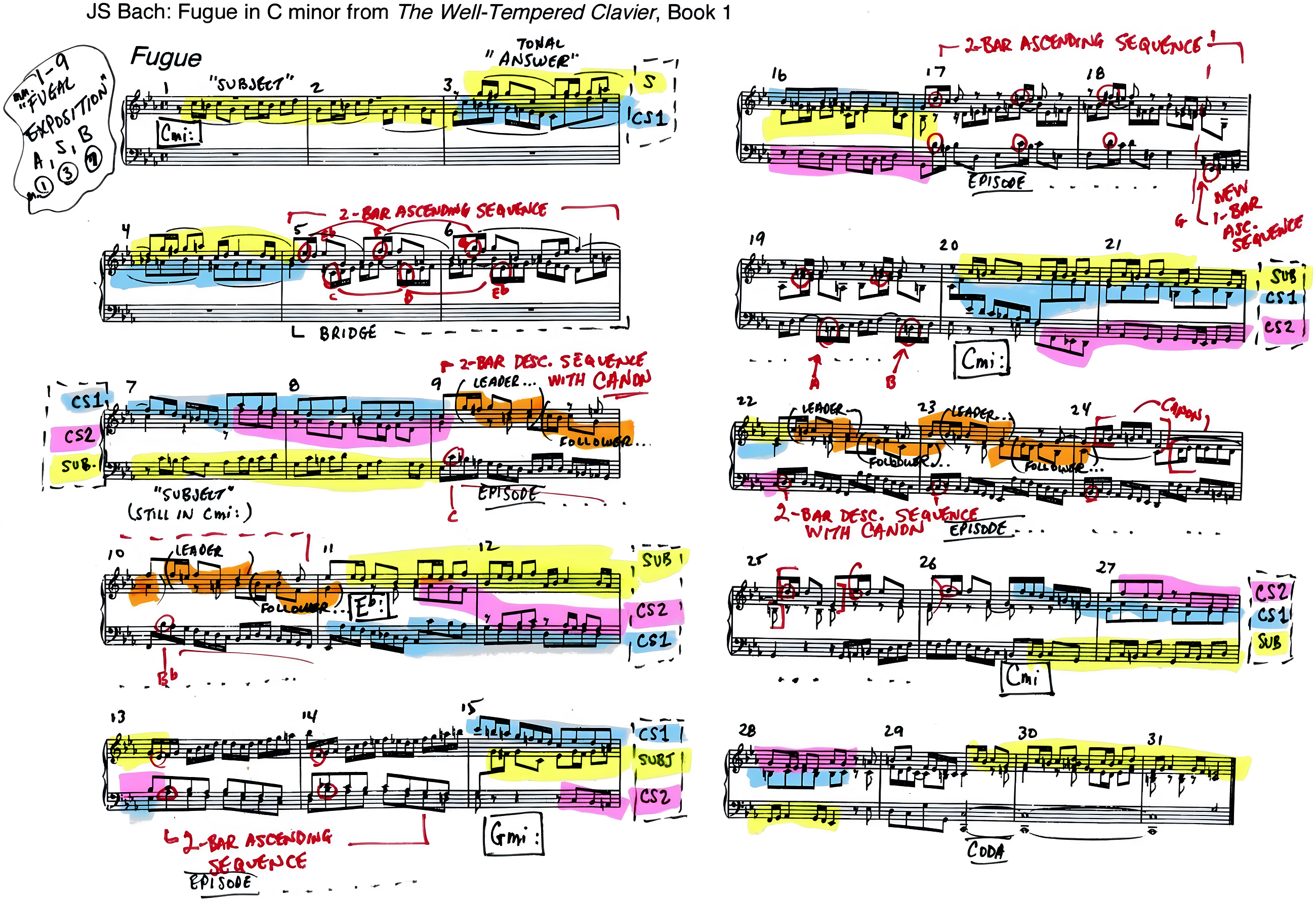}
    }
    \vspace*{-12pt}
    \caption{
    This manual fugue analysis example shows how a music analyst would perform a melodic pattern analysis using color and annotations.
    We took this example as a solution from an exam published by the School of Music from Western Michigan University~\cite{wmich2019fuguereview}.    
    }
    \label{fig:wmich2019fuguereview}
\end{figure}

We cooperated with domain experts and musicologists to extract requirements and tasks regarding melodic analysis.
In addition, we consider an exemplary Fugue analysis provided by the School of Music from Western Michigan University (see~\autoref{fig:wmich2019fuguereview}) revealing how manual melodic analysis is performed, which accords with the expert input.

\newcommand{\reqA}{\texttt{R1}}
\newcommand{\reqB}{\texttt{R2}}
\newcommand{\reqC}{\texttt{R3}}
\newcommand{\reqD}{\texttt{R4}}
\newcommand{\reqE}{\texttt{R5}}
\newcommand{\reqF}{\texttt{R6}}

\newcommand{\reqAP}{\texttt{[\reqA]}}
\newcommand{\reqBP}{\texttt{[\reqB]}}
\newcommand{\reqCP}{\texttt{[\reqC]}}
\newcommand{\reqDP}{\texttt{[\reqD]}}
\newcommand{\reqEP}{\texttt{[\reqE]}}
\newcommand{\reqFP}{\texttt{[\reqF]}}

Drawing from this analysis example, we can identify four key requirements that are fundamental to supporting music analysts in their exploration of melodic patterns and their transformations.

\subhead{\reqA: Identify relevant melodic sequences --} An analyst must be able to identify and select melodic motifs. 
It is relevant only to consider notes from the same voice.
Thus, constraining patterns to remain within a single voice is crucial. 
This pattern identification is typically an in-situ annotation using the Common Music Notation~\cite{Schottstaedt1997CMN} which is the representation analysts are most familiar with.

\subhead{\reqB: Detect melodic pattern reoccurrences across the piece in different voices --} Identifying reoccurrences and variations of a theme typically requires analyzing multiple voices separately to determine how a composer develops a melodic topic during a composition. 
Thus, a visual analysis system must be capable of locating and highlighting these melodic repetitions across different voices, facilitating a comprehensive understanding of a piece's polyphonic texture.

\subhead{\reqC: Identify sophisticated variations and reoccurrences of a given pattern --} 
The handwritten analysis also reveals instances where the original melodic patterns are transformed through operators such as inversion~\opB or transposition~\opA.
Therefore, it is relevant to automatically recognize and display these modified patterns throughout a piece, enabling the analyst to track the evolution of melodic themes.

\subhead{\reqD: Interpret the meaning of the pattern, its variation type, and its position in the piece --}
The exemplary analysis provides insights into the musical significance and context of the identified patterns and variations. 
\toolname should integrate techniques that help understand the meaning and implications of the detected patterns, including the type of variation applied and their role in the larger composition context.

\newbool{usewidefigure}
\booltrue{usewidefigure}

\newcommand{\SV}{\texttt{SV}\xspace} 
\newcommand{\MOS}{\texttt{MOS}\xspace} 
\newcommand{\VSV}{\texttt{VSV}\xspace} 
\newcommand{\MTG}{\texttt{MTG}\xspace} 
\newcommand{\MPC}{\texttt{MPC}\xspace} 

\section{MelodyVis -- Visual Interface for Melody Analysis} 
\noindent \toolname is a visual analytics (VA) approach to melodic pattern analysis of sheet music, which is intended for music analysts. 
Below, we provide conceptual details about the interconnected components of the application and introduce the required processing steps.
We also propose directions on how our approach could be further improved.

\subsection{Data Preparation}

Analyzing a music sheet in \toolname requires initial pre-processing steps:
First, we require that any composition must be available using the MusicXML format, a widely used, general, and flexible format to encode symbolic music information~\cite{good_musicxml_2001}.
We leverage the music21 Python library~\cite{cuthbert_music21_2010} to load the digital sheet music files, continuing with the data processing.
To identify coherent voices, we leverage the envelope extraction technique also used by Gray and Bunescu~\cite{gray2020fromnotelevel} that allows retrieving the separate voices from a composition.
While a general data importer is not available at this point, it would be easy to extend it, as we loop through a predefined set of compositions that are currently available in \toolname.

\subsection{Interaction Components}

The visual interface for analyzing melodic patterns based on digital sheet music comprises five interconnected core components supporting the interactive analysis workflow as indicated in~\autoref{fig:teaser}: (1) Sheet View, (2) Melodic Operator Selection, (3) Voice Separation Overview, (4) Melody Transformation Graph, and (5) Melodic Pattern Configuration.
The five separate components are updated based on the user interactions that analysts typically use in the same order as we describe them in the following.

\subhead{Sheet Music View}

\newcommand{\sheetviewwrapfig}{
\wrapfigcustom{15}{r}{4cm}{4cm}{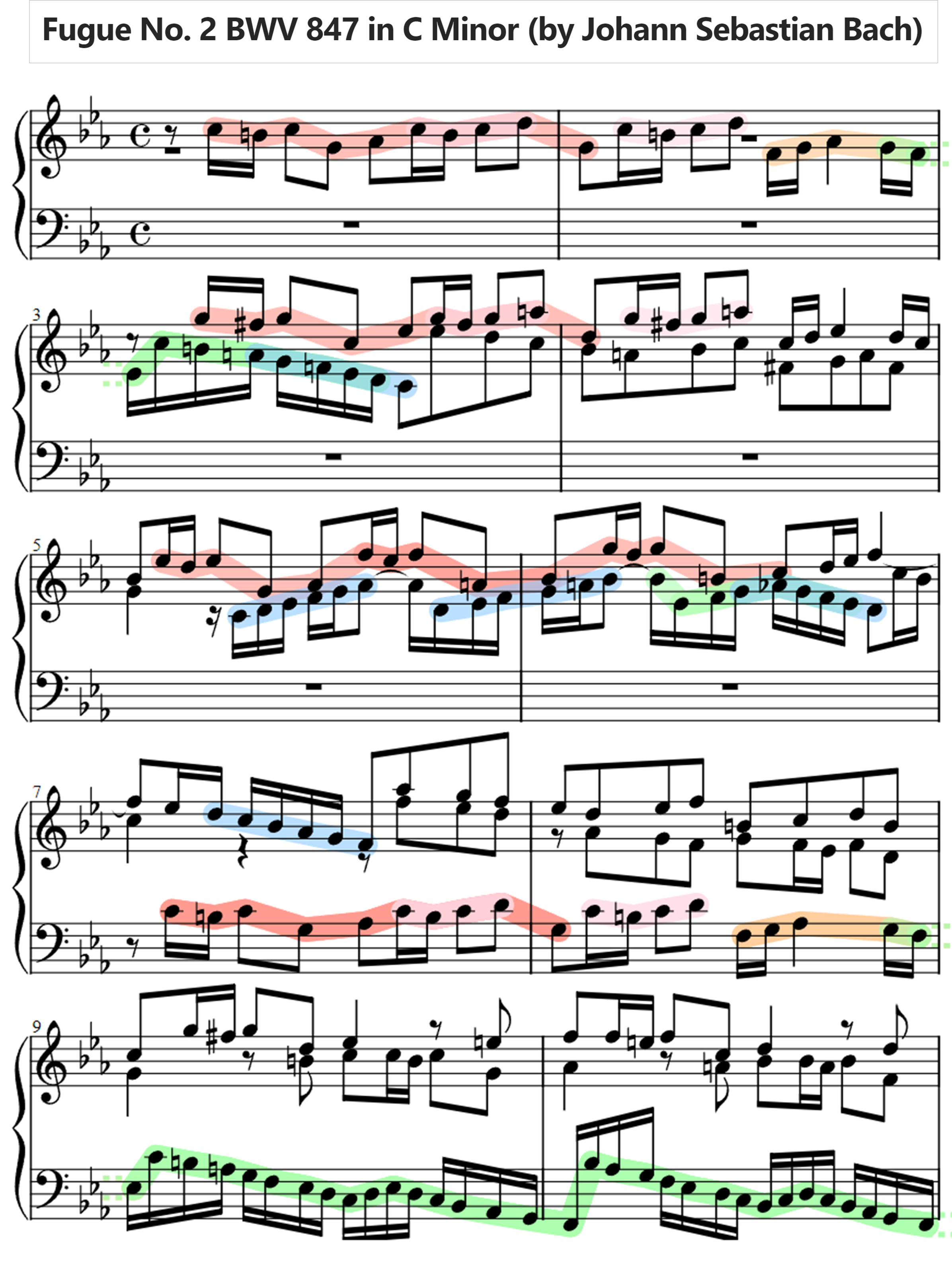}{-35pt}{-5pt}
}
\ifbool{usewidefigure}{}{\sheetviewwrapfig}
\noindent The sheet view (\SV) facilitates the easy selection of initial melodic patterns based on the separated voices~\reqAP.
This intuitive interface enables analysts to configure note coloring based on voices, facilitating visual differentiation as needed. 
The \SV features in-situ placement of detected patterns by \toolname, utilizing different colors to distinguish between variations~\reqCP. 
Furthermore, the component automatically adjusts measure splitting according to the analyst's screen size, optimizing readability and user experience.

\ifbool{usewidefigure}{
\begin{figure*}[t]
    \centering
    \href{http://visual-musicology.com/melodyvis/}{\includegraphics[width=\linewidth]{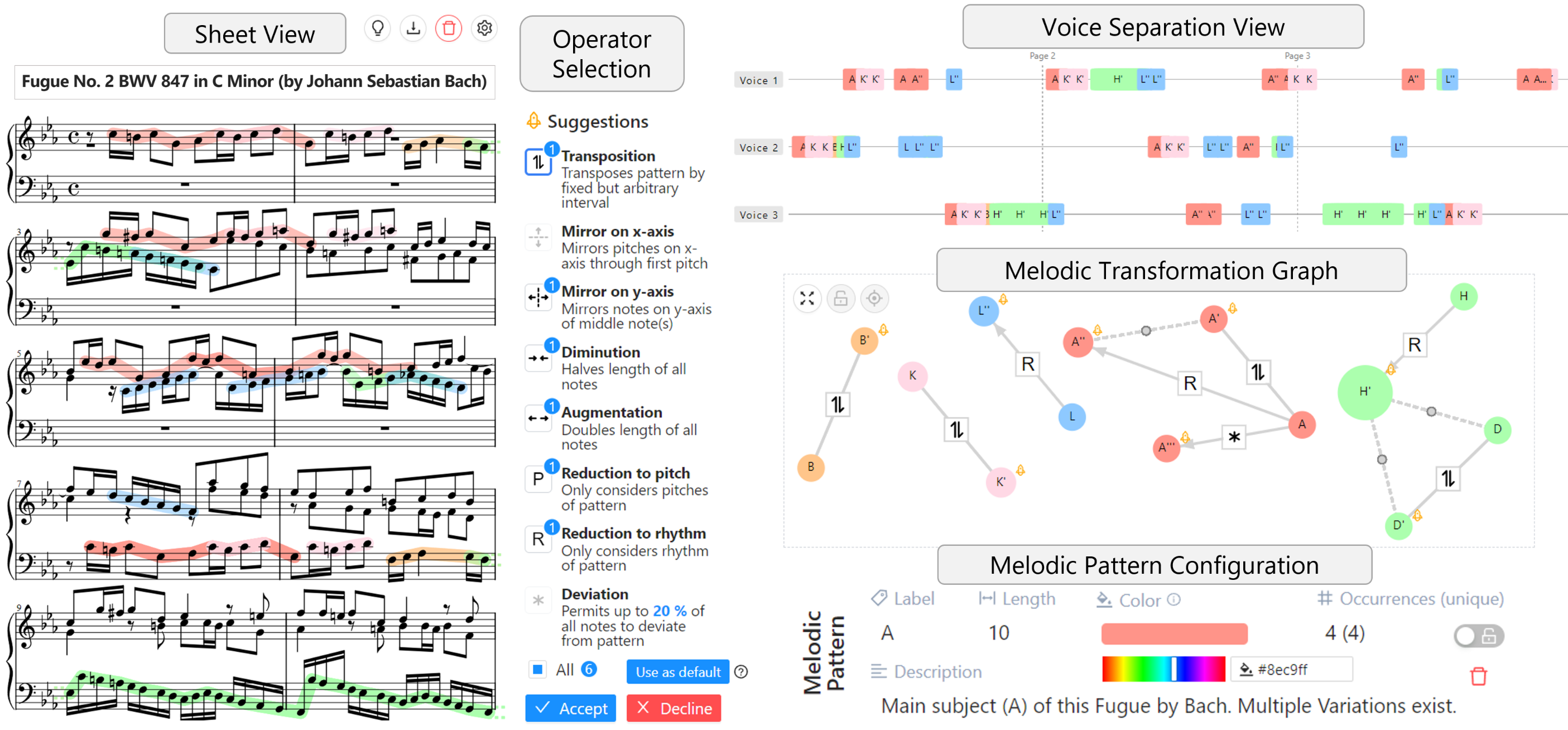}}    
    \caption{ 
    \toolname consists of five core interaction components interconnected through linking and brushing~\cite{keim2002infovisanddatamining}:
     (1) Sheet Music View, (2) Melodic Operator Selection, (3) Voice Separation View, (4) Melodic Transformation Graph, and (5) Melodic Pattern Configuration. 
     Together, they facilitate a comprehensive and interactive analysis workflow for the exploration of melodic patterns in digital sheet music.
    }
    \label{fig:melodyvis_app}
\end{figure*}
}{}

\FloatBarrier
\noindent
Combining these features allows the \SV to provide a suitable starting point for melody analysis while highlighting identified patterns and their variations.
As the first step in the analysis, the \SV highlights the subsequent notes from the same voice when selecting an initial note, assisting users in completing their pattern selection.

\subhead{Melodic Operator Selection}
As shown in~\autoref{fig:teaser}, the introduced atomic melodic operators are applied to the initial melodic patterns selected within the sheet view. 
The application of these operators generates a list of pattern variations, which allows users to further explore the melodic patterns by chaining multiple operators together. 
This Melodic Operator Selection (\MOS) enables creating more advanced, complex transformation chains, ultimately facilitating an efficient expansion of the search space during melodic pattern investigations.

\ifbool{usewidefigure}{}{
\begin{figure}[ht]
    \centering
    \addbordertofigure{\includegraphics[width=\linewidth]{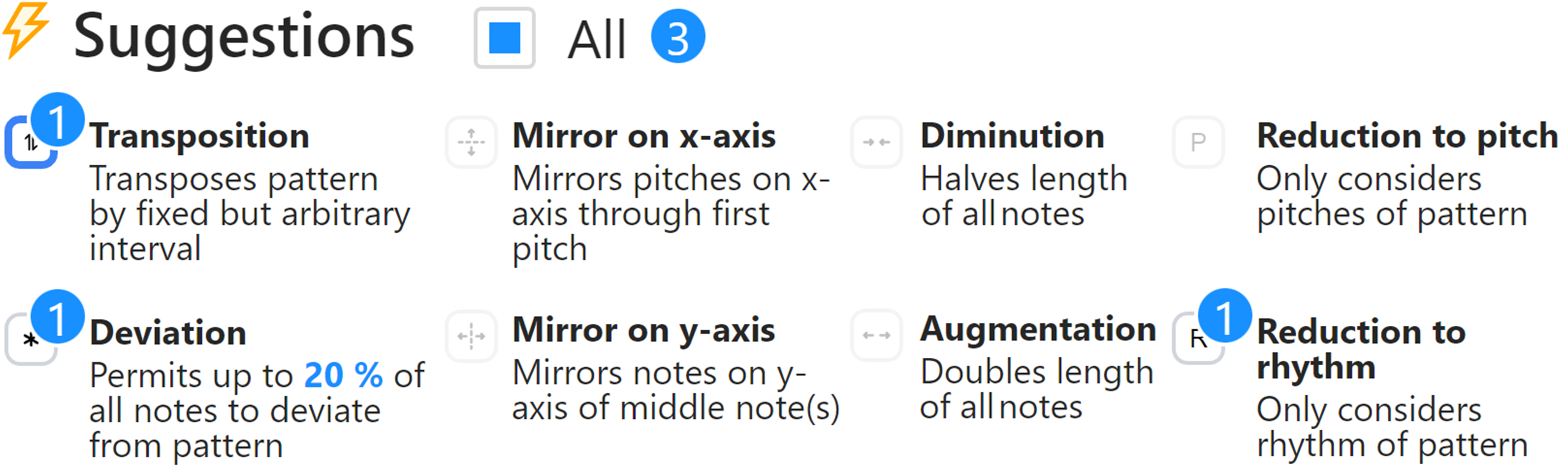}}
    \caption{
    Upon selecting a melodic pattern, the atomic melodic transformation operators are highlighted when the prototype identifies variations using a specific operator.
    }
    \label{fig:melodyvis_atomic_operators}
\end{figure}
}

\subhead{Voice Separation View}
\noindent
The \emph{Voice Separation View} (\VSV) provides a comprehensive overview of detected patterns throughout the sheet.
Page markers as vertical lines are a navigational aid that show at which positions the page breaks occur in the \SV.
The mouse position in the \VSV shows the corresponding page in the \SV.
This view effectively displays the distribution of patterns across all voices within a given sheet, using consistent coloring as in the sheet view.
Users can access additional detailed information about the patterns when hovering over elements. 
If required, the \VSV enables melody analysts to select patterns for further annotation editing. 
With the appropriate coloring settings applied, analysts can easily identify the first occurrence of a specific pattern within a particular voice, streamlining the analysis.

\ifbool{usewidefigure}{}{
\begin{figure}[ht]
    \centering
    \addbordertofigure{\includegraphics[width=\linewidth]{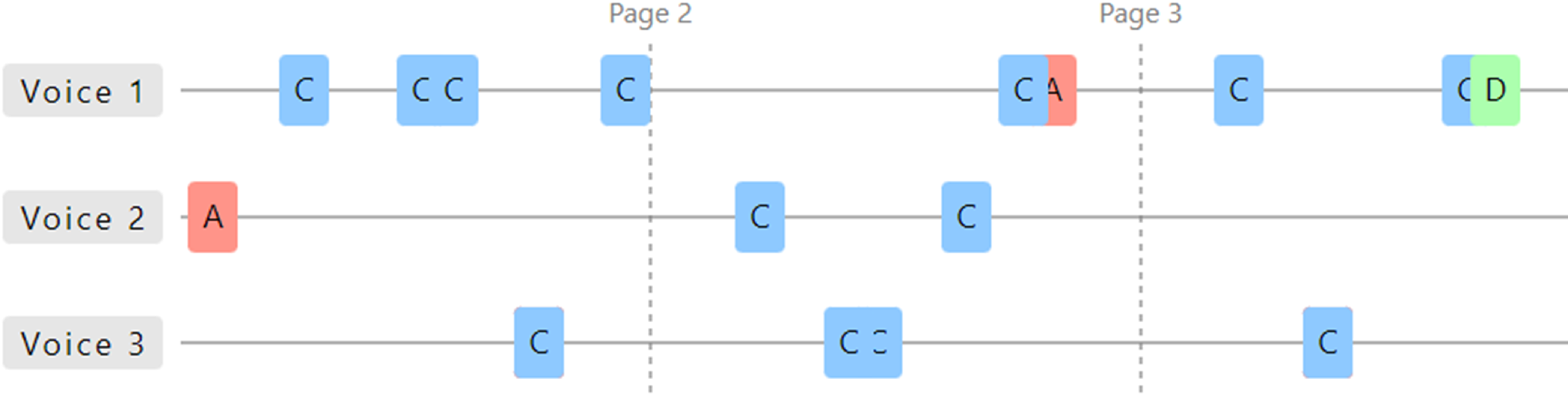}}   
    \caption{
    The \emph{separate voice view} helps to compare the occurrence of musical patterns between different voices.
    }
    \label{melodyvis_voice_view}
\end{figure}
}

\subhead{Melodic Transformation Graph}

\noindent The Melodic Transformation Graph (\MTG) is a supportive and more abstract visualization component designed to display the transitions of a melodic pattern selection to a set of other patterns resulting from applying a specific operator~(see~\autoref{fig:melodyvis_app}). 
Nodes within the graph represent either single patterns when selected or entire pattern groups based on an applied operator, with the direction of transformation indicated by representative arrows. 
Operator icons situated on the edges between nodes facilitate the identification of which operator was used to obtain the resulting pattern set.

Connected graph components reveal related patterns, while dashed lines depict similar patterns in different result sets when applying distinct operators, thus highlighting similarities between operators. 
This comprehensive visualization enables analysts to gain deeper insights into the complex relationships between melodic patterns and the effects of applied transformation operators. 
Analysts can explore the intricate connections between melodic patterns and their variations within a given musical composition through the \MTG.
This components also helps to differentiate between user-selected melodic patterns and automatic detection that correspond to a prior user selection.

\ifbool{usewidefigure}{}{
\begin{figure}[ht]
    \centering
    \addbordertofigure{\includegraphics[width=\linewidth]{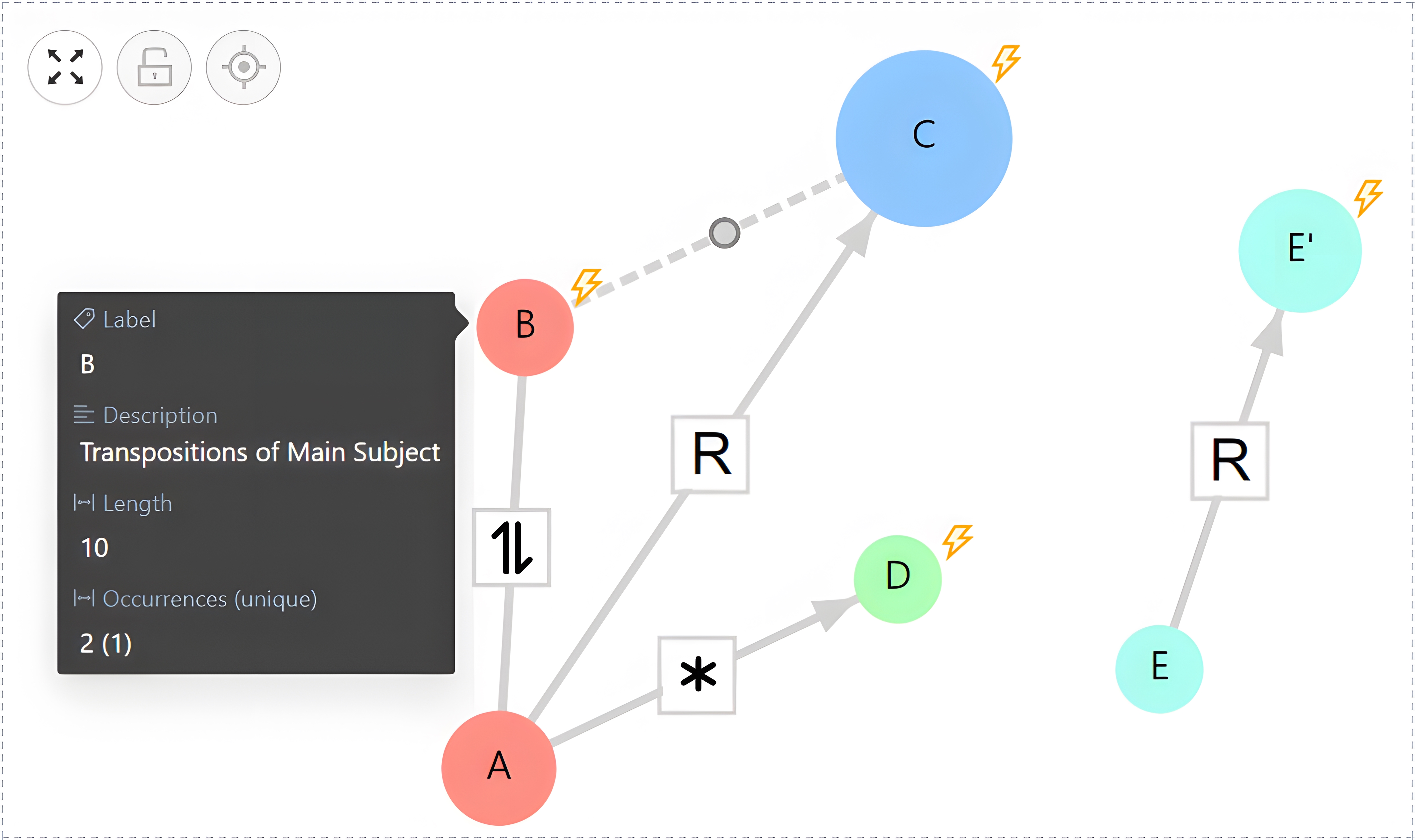}
    }
    
    \caption{
    Showing transitions between sets of melodic patterns found when using an atomic operator.
    }
    \label{melodyvis_operator_graph}
\end{figure}
}

\subhead{Melodic Pattern Configuration --} 
\noindent The Melodic Pattern Configuration (\MPC) is a versatile component that provides users with additional details and customization options when they click on a melodic pattern within any other components (\SV, \VSV, \MTG). 
This dedicated component displays summarized information about a pattern, such as unique or overlapping occurrences and pattern length (i.e., number of notes).

Analysts can tailor short labels to suit their needs and modify the coloring of a single pattern or a pattern group using a simple color map. 
While \toolname primarily uses distinct pastel colors to emphasize different patterns, users can still adjust colors to represent similarities according to their preferences. 
In the future, the coloring could be adapted based on the similarity space of two patterns.
The component also allows for adding a detailed description (up to 280 characters) for knowledge externalization~\reqDP. 
Also, a deletion button is provided, enabling users to remove single patterns as needed. 
This comprehensive configuration component facilitates a more customized and efficient analysis experience for music analysts.
The manually changed annotations are linked via tooltips to the other components such that they can easily be reused during the analysis and for comparing multiple melodic patterns.

\ifbool{usewidefigure}{}{
\begin{figure}[ht]
    \centering
    \addbordertofigure{\includegraphics[width=\linewidth]{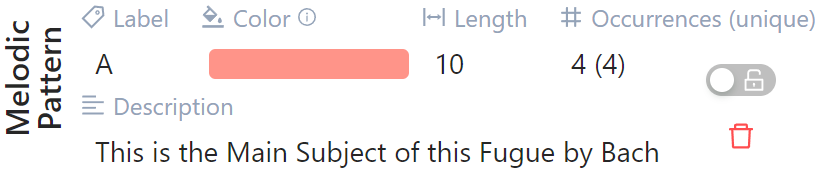}}   
    \caption{
    Number of occurrences, change color.. add label add description and total length of current pattern.
    }
    \label{melodyvis_pattern_details}
\end{figure}
}

\subsection{Visual Interaction Pipeline} 
\noindent 
\toolname follows a visual interaction pipeline designed to facilitate the interactive analysis of melodic patterns in sheet music. 
As described in~\autoref{fig:teaser}, analysts can begin by selecting an initial melodic pattern within the \SV, which is then utilized to apply melodic operators that generate further pattern highlights. 
Following this, the \VSV provides a comprehensive overview and distribution of the different voices present in the composition. 
To support the analysis of melodic variations, the \MTG visually displays the relationships between the initial pattern and its transformed counterparts. 
As the analyst progresses, they can seamlessly loop through this analysis process to continue discovering and exploring additional melodic patterns within the sheet music.

\section{Use Cases}

Subsequently, we will present three distinct use cases to demonstrate the interactive analysis workflow with \toolname and highlight its capabilities in analyzing digital sheet music for melodic patterns. 
With these use cases, we aim to showcase our prototype's practical applications and provide a solid understanding of how users can leverage \toolname to gain insights into musical compositions, emphasizing its potential in both educational and research contexts.

\subsection{The Four Seasons: Spring -- by Antonio Vivaldi}

\begin{figure}[h]
    \vspace*{-5pt}
    \centering
    \addbordertofigure{
    \includegraphics[width=0.96\linewidth]{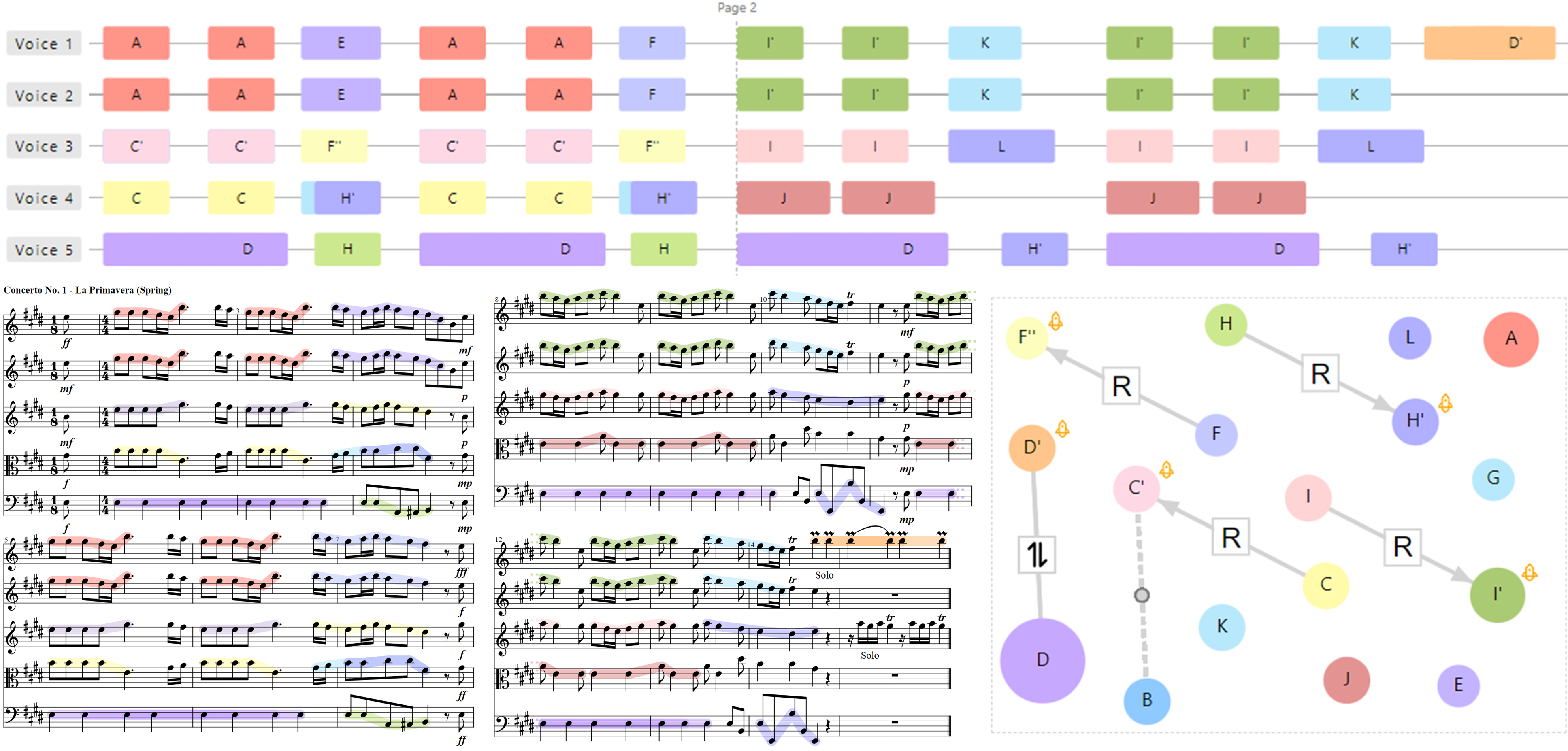} 
    }
    \vspace*{-12pt}
    \caption{ 
     Visual analysis of Vivaldi's \href{https://musescore.com/musikmann/scores/135805}{1\tss{st} movement of ``The Four Seasons - Spring''}~\cite{vivaldi1680spring}. 
     (Top) \VSV with color-coded highlights indicating corresponding melodic patterns across the voices. 
     (Bottom Left) The \SV displays the selected patterns with overlay annotations. 
     (Bottom Right) The MTG illustrates the relationships between the identified patterns and their transformations within the composition showing dashed edges with pattern sets that have similar patterns.
    }
    \label{fig:usecase:vivaldi_spring}
\end{figure}

The first use case demonstrates how a user can explore and identify melodic patterns in Vivaldi's ``The Four Seasons - Spring (La Primavera)'' 1\tss{st} movement~\cite{vivaldi1680spring}. 
The composition features five distinct voices: three violin parts, one viola, and one violoncello part as depicted in the \VSV at the top in~\autoref{fig:usecase:vivaldi_spring}.

\noindent First, the user selects the second note from the first voice (Violin 1) directly on the sheet music (\SV) as the starting note of the first pattern. 
A second note from the same voice must be selected to complete the first pattern~\reqAP.
The user can use this interactive exploration process to identify similar occurrences of this first pattern (light red) in the second voice~\reqBP.
Next, the user examines the melodic variations in the 3rd and 4th voices (Violin 2 and Viola), which play along with a similar pitch~\opF. 
By doing so, the user can observe the aligned similarities between these voices.
In the fifth voice (Violoncello), the user can see how it lays a solid foundation by playing longer notes with lower pitches and repeating the melodic texture. 
The Voice Separation View proves to be particularly helpful in identifying the similarities between the corresponding voices and the regular structure of the composition.
The \MTG aids the analyst in visualizing the connections~\reqCP~between exact patterns and instances where the main themes are reduced to rhythm only~\opG. 
Finally, in the \MPC, the analyst adds a short description about the main theme that Vivaldi creates a vivid representation of the Spring season's essence~\reqDP.

\subsection{BWV 846: Prelude in C Major -- by J. S. Bach}

\begin{figure}[ht]
    \vspace*{-5pt}
    \centering
    \addbordertofigure{
    \includegraphics[width=0.96\linewidth]{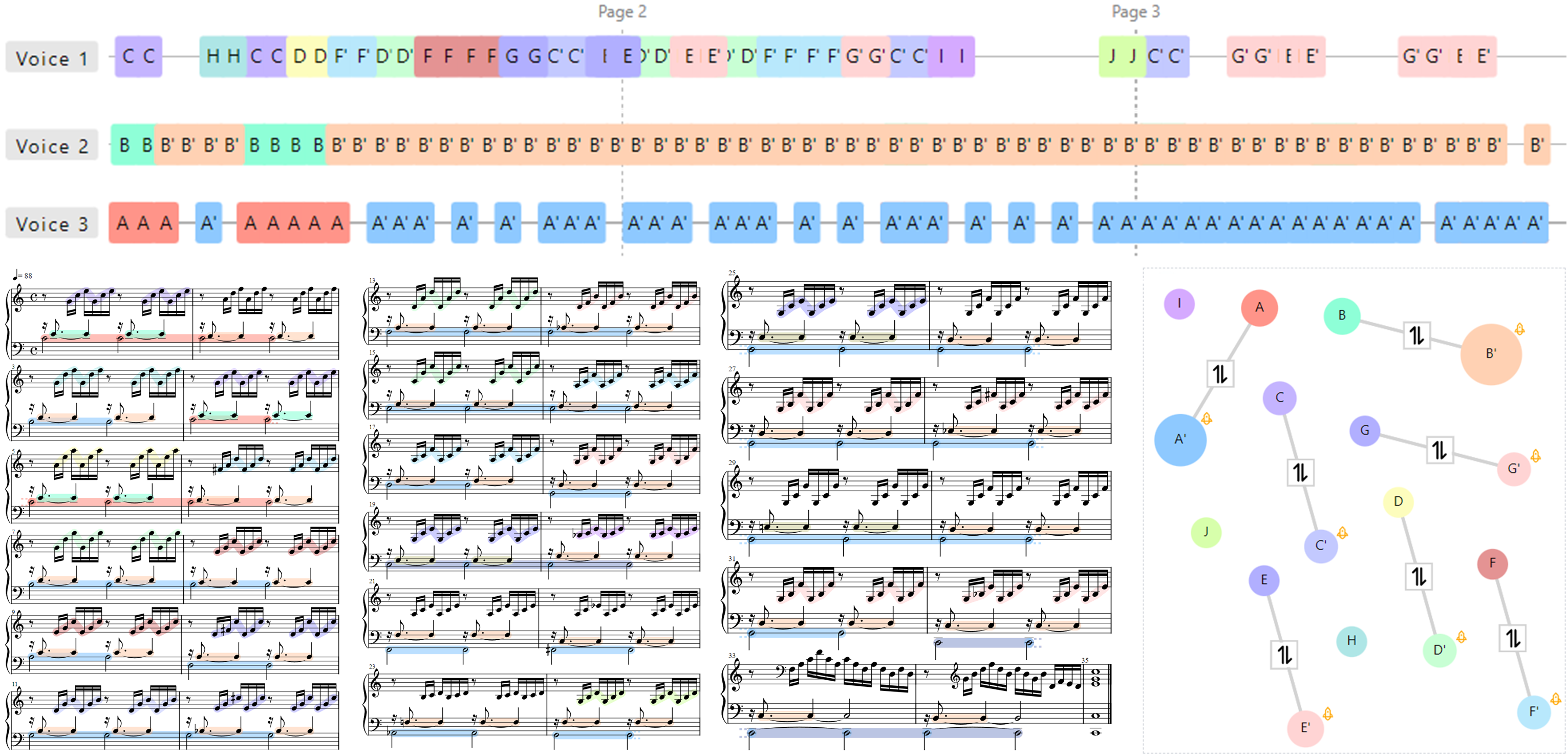} 
    }
    \vspace*{-12pt}
    \caption{ 
      Analysis of J.S. Bach's \href{https://musescore.com/user/101554/scores/117279}{``Prelude in C Major (BWV 846)''}, focusing on the arpeggio patterns in the first voice, where the most interesting melodic development occurs, while illustrating the simple repetitions occurring in the second and third voice.
    }
    \label{fig:usecase:bach_preludeinc}
    \vspace*{-5pt}
\end{figure}

In this second use case, we examine Bach's \href{https://musescore.com/user/101554/scores/117279}{``Prelude in C-Major (BWV 846)''}~\cite{bach1700preludeincmaj}. 
This piece is characterized by its upward-directed melodic line and distinctive broken chord patterns. 
The analysis begins by selecting an upward arpeggio pattern in the first voice. 
Using the \VSV and \SV, the analyst can then view all transposed matches of the same pattern in the same voice~(see~\autoref{fig:usecase:bach_preludeinc}).

The \VSV emphasizes the simple repetitions of the same pattern in the second and third voices throughout the piece, while the most melodic changes are present in the first voice.
By showcasing how \toolname can effectively analyze and illustrate transposition and upward broken chord patterns (arpeggios) in a well-known composition like BWV 846, this use case demonstrates the tool's potential for enhancing the understanding of repetitive melodic structures and relationships.

\noindent
The second and third voices comprise only two transposed patterns. 
While musicological experts can identify uniformity by examining sheet music, precise recurrences and transpositions in arpeggios are easier to discern through linking and brushing on the \VSV.

The contrast between the simplicity of the second and third voices and the more varied first voice showcases \toolname's ability to reveal varying complexity within a single composition. 
Our prototype is versatile enough to handle diverse melodic structures through integrated atomic operators independent of underlying complexity. 
This contrast emphasizes the importance of visual techniques in providing traceable, detailed analysis of melodic patterns and relationships, ultimately enhancing music understanding.

\subsection{Tetris Theme -- by Nikolai Alexejewitsch Nekrassow}

The third use case explores a single-voice version of the Tetris theme by Nikolai Alexejewitsch Nekrassow. 

\begin{figure}[b]
    \vspace*{-5pt}
    \centering
    \addbordertofigure{
    \includegraphics[width=0.96\linewidth]{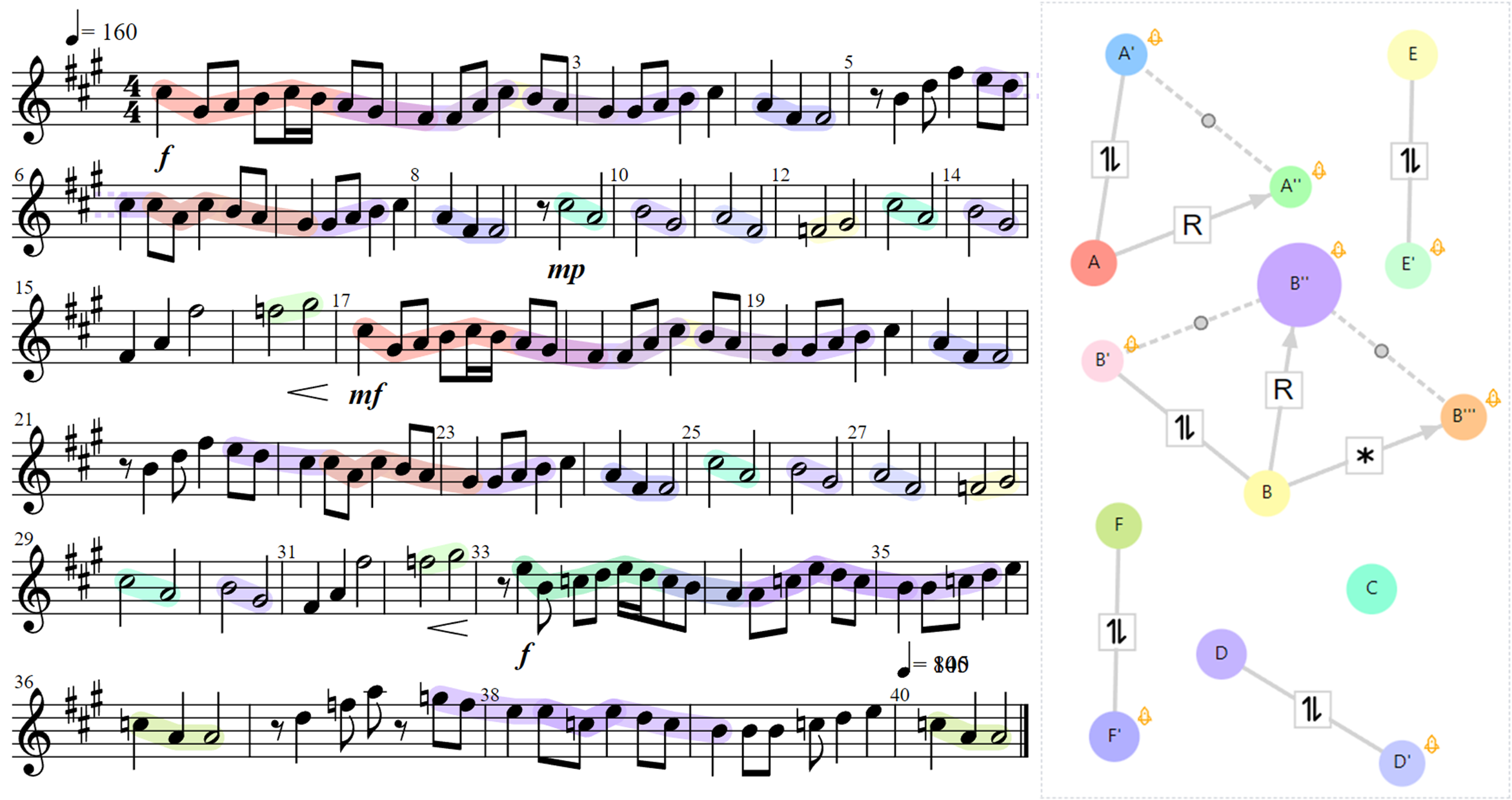} 
    }
    \vspace*{-12pt}
    \caption{ 
      Tetris theme analysis: On the left, the highlighted Sheet View displays the recurring motifs and contrasting melodic sequences. 
      The Melodic Transformation Graph on the right reveals the simple relationships yet memorable patterns used in this iconic melody.
    }
    \label{fig:usecase:tetris}
\end{figure}

The analysis begins with identifying the starting theme~\reqAP, which occurs three times (highlighted in light red in~\autoref{fig:usecase:tetris}). 
The MTG displays the theme's relationships, allowing users to visualize the abstract structure of the melody patterns and their recurring motifs~\reqBP. 
Some edges between the graph nodes are drawn using dashed lines in specific scenarios.
These dashed lines identify bridges between pattern sets resulting from using the operators~\reqCP.
Tooltips on demand indicate how many similar patterns occur on both sides of the edge.
The \MTG also reveals independent graph components that emphasize connected melodic pattern groups, providing the provenance of all hitherto identified patterns.
In contrast to the main theme, in bars 9-11 the downward half notes using minor and major thirds provide a contrasting melodic sequence, adding further, even though very simple, variety.

\toolname helps gain insights into this well-known theme~\reqDP, illustrating its applicability for analyzing various musical compositions, from classical works to more straightforward, single-voice melodies.

\begin{figure*}[t]
	\includegraphics[width=\linewidth]{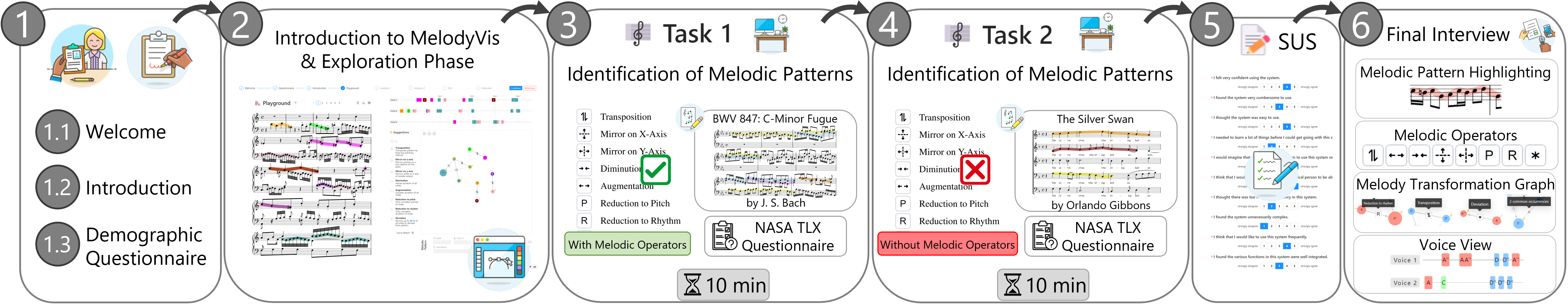}
 \caption{User Study Procedure --
 (1) Welcome, introduction video, demographic questionnaire, and textual explanation of melodic operators; 
 (2) Familiarization with the prototype; 
 (3) First task, followed by NASA-TLX workload evaluation; 
 (4) Second task with randomized music piece and operator availability; 
 (5) System Usability Scale (SUS); 
 (6) Final survey with an interaction questionnaire about the different components.}
 \label{fig:studydesign}
\end{figure*}

\FloatBarrier

\section{Online User Study}
\noindent We conducted an online user study to assess our \toolname prototype in facilitating melody analysis tasks for participants with diverse backgrounds and expertise levels.

\newcommand{\smA}{\texttt{\href{https://musescore.com/classicman/scores/231261}{SM1}}}
\newcommand{\smB}{\texttt{\href{https://musescore.com/user/33131263/scores/5783924}{SM2}}}

\newcommand{\taskwithop}{\texttt{$T_w$}}
\newcommand{\taskwithoutop}{\texttt{$T_{wo}$}}

\subsection{User Study Design}  

\subhead{Datasets --}
\noindent In the study, we utilized two distinct music sheets, carefully chosen to include multiple independent voicings. 
The first music sheet \smA~is \href{https://musescore.com/classicman/scores/231261}{\emph{``Fugue No. 2 in C Minor (BWV 847)''} by Johann Sebastian Bach}, a well-known Baroque piece characterized by its intricate counterpoint and elaborate melodic structures~\cite{vonbachFugueNoBWV847}. 
The second music sheet \smB~is \href{https://musescore.com/user/33131263/scores/5783924}{\emph{``The Silver Swan''} by Orlando Gibbons}~\cite{gibbonsSilverSwan}.
This early 17th-century English madrigal comprises a different style and compositional approach, with rich textures and expressive melodies. 
These two pieces provide participants with varied challenges and opportunities to explore the capabilities of our prototype while engaging with different musical styles and amount of voicings.

\newcommand{\p}[2]{{$\mathbf{#1}_\mathbf{#2}$}}
\newcommand{\participantA}{\p{P}{1}}  
\newcommand{\participantB}{\p{P}{2}}  
\newcommand{\participantC}{\p{P}{3}}  
\newcommand{\participantD}{\p{P}{4}}  
\newcommand{\participantE}{\p{P}{5}}  
\newcommand{\participantF}{\p{P}{6}}  
\newcommand{\participantG}{\p{P}{7}}  
\newcommand{\participantH}{\p{P}{8}}  
\newcommand{\participantI}{\p{P}{9}}  
\newcommand{\participantJ}{\p{P}{10}} 
\newcommand{\participantK}{\p{P}{11}} 
\newcommand{\participantL}{\p{P}{12}} 
\newcommand{\participantM}{\p{P}{13}} 
\newcommand{\participantN}{\p{P}{14}} 
\newcommand{\participantO}{\p{P}{15}} 
\newcommand{\participantP}{\p{P}{16}} 
\newcommand{\participantQ}{\p{P}{17}} 
\newcommand{\participantR}{\p{P}{18}} 
\newcommand{\participantS}{\p{P}{19}} 
\newcommand{\participantT}{\p{P}{20}}

\subhead{Participants --} 
\noindent
We recruited 25 participants of different expertise levels through various channels, including e-mail and online platforms such as 
Discord, LinkedIn, Reddit, and Twitter. 
Our e-mail recruitment primarily targeted mailing lists for professional musicians and musicologists, music associations, members of philharmonic orchestras, and selected individuals with a keen interest in music. 
On Discord and Reddit, we posted calls for study participation in communities that bring together hobby musicians and musicologists. 
This multi-channel recruitment strategy allowed us to gather a heterogeneous group, encompassing different backgrounds, experiences, and levels of expertise in music theory, melody analysis, and data visualization.

\subhead{Study Tasks --}
\noindent
The participants were tasked with identifying melodic patterns they deemed relevant in a given music sheet by highlighting a melodic pattern (i.e., selecting the first and the last note).
To measure the impact of the melodic operators, the study involved \textbf{two rounds in randomized order} to minimize learning effects: analyze one of the music sheets (\smA~and \smB) once \emph{with} and once \emph{without} the melodic operators available leading to a total of four different trial setups.
For the former, we also instructed the participants to use the available melodic operators to identify melodic patterns.
They were advised to focus on identifying longer melodic patterns in favor of shorter ones, particularly those that would stand out to a listener of this music sheet. 
Participants were expected to give meaningful labels and descriptions of their identified melodic patterns. 
There was no particular solution to this task; hence, the participant could complete the task as soon as they considered it completed. 
We especially considered the amount and type of patterns
Participants could expect to identify between 10 and 20 melodic patterns and had 10 minutes to complete each round of tasks.

\subhead{Study Procedure --}
\noindent 
When our participants opened our web-based study prototype (\href{https://visual-musicology.com/melodyvis-study/}{\texttt{visual-musicology.com/melodyvis-study}}), they were provided with a short textual introduction to the study procedure and prompted to agree to the terms of the study.
Our study design to assess \toolname is described in detail in~\autoref{fig:studydesign}. 
We used Bach's ``\href{https://musescore.com/classicman/scores/1777656}{Prelude in A minor, BWV 889}'' in the exploratory phase to let the participants get acquainted with the components~\cite{vonbachFugueNoBWV889}.

\subsection{Study Results} 
\begin{figure}[b]
    \vspace*{-7pt}
    \centering
    \addbordertofigure{\includegraphics[width=0.96\linewidth]{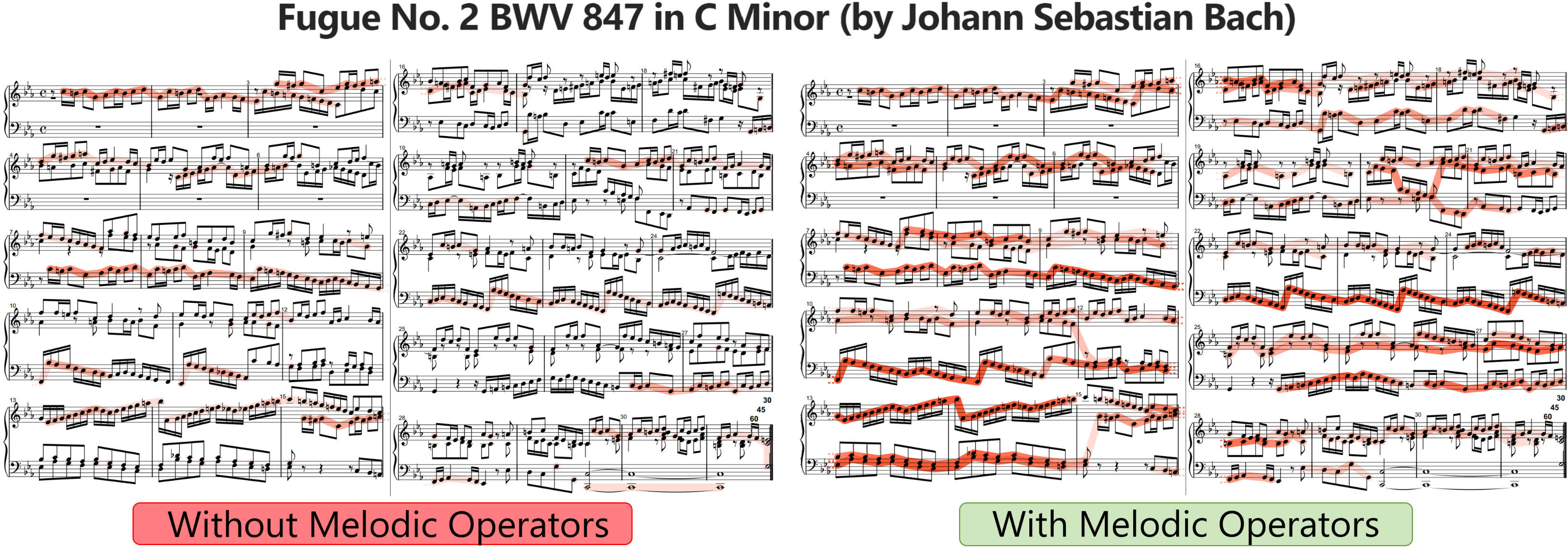} }
    \vspace*{-12pt}
    \caption{ 
    Resulting patterns on \smA~for both conditions. 
    The two music sheets reveal that with the operators (right), at least twice as many patterns could be identified compared to those without (left).
    }
    \label{fig:evaluation:cond_results_fugue}
\end{figure}

\subhead{Qualitative Results --} 
In a final survey using an interaction questionnaire we gathered positive and negative feedback from our participants.
Initially, we excluded all participants from the study who did not provide any patterns or would not complete the whole study.

\noindent 
Several participants emphasized our approach's benefits, including its ability to classify different operators and \textbf{assist} in analyzing and improving sheet music (\participantA). 
The system's \textbf{practicality} and intuitiveness were praised, with users finding the melody highlighting and timeline features particularly helpful (\participantB, \participantS, \participantT). 
The \textbf{speed and efficiency} of \toolname in identifying patterns were appreciated by participants (\participantD, \participantR). 
Furthermore, the visually \textbf{seamless interface} contributed to a positive user experience (\participantE, \participantP). 
Other valued features included the ease of \textbf{marking melody patterns} (\participantM, \participantO), the \textbf{automatic recognition} of recurring motifs (\participantL), and the \textbf{suggestions} of melodic operators (\participantF, \participantG, \participantH). 
Overall, the graphical representation on sheet music was well received (\participantN), and the highlighting of melody components (\participantT).

\noindent A few participants also raised concerns regarding their experience with \toolname. 
Some encountered \textbf{difficulties selecting notes}, as the highlight would jump to other staves (\participantE). 
The lack of an \textbf{auditorial reference} made it challenging for some users to analyze the melodic patterns (\participantS), especially without the operators in place. 
A few users reported that the application failed to recognize specific melodic patterns due to \textbf{chromatic variations} (\participantP). 
Additionally, some participants were\textbf{ unclear} about the \textbf{goals} of the tasks (\participantJ) or found navigating the interface to have a steep learning curve (\participantF).

\subhead{Task-solving Strategies --}
From the final questionnaire, we identified multiple strategies the participants employed to identify patterns during their tasks. 
They relied on (1) \emph{familiarity and repetition} by recognizing recurring elements and focusing on frequently appearing motifs. 
Users also paid attention to (2) \emph{musical characteristics}, such as large leaps, unusual rhythms, fugal subjects, and imitated fragments to guide their analysis.
They considered the (3) \emph{context and structure} of the music, looking for patterns around the beginnings of sections, voice entrances, lyrics, or changes in the composition's character. Lastly, participants used their (4) \emph{intuition and prior knowledge}, imagining the melody in their heads or trying sight-singing to better understand the music and identify related patterns. 
By understanding these strategies, we can refine \toolname to better support users' melodic analysis efforts.

\subhead{Results of Selected Patterns --} 
When comparing the selected patterns of both compositions used in the study, we identified that the users could detect at least twice as many patterns when using the melodic operators. \autoref{fig:evaluation:cond_results_fugue} depicts the music sheet for \smA~indicating with a red highlighting the distribution of all found patterns. 
While some participants found patterns without the operators which others did not, we identified that most of the participants were able to identify the same patterns with the operators.

\begin{figure}[h]
    \vspace*{-7pt}
    \centering
    \addbordertofigure{\includegraphics[width=0.96\linewidth]{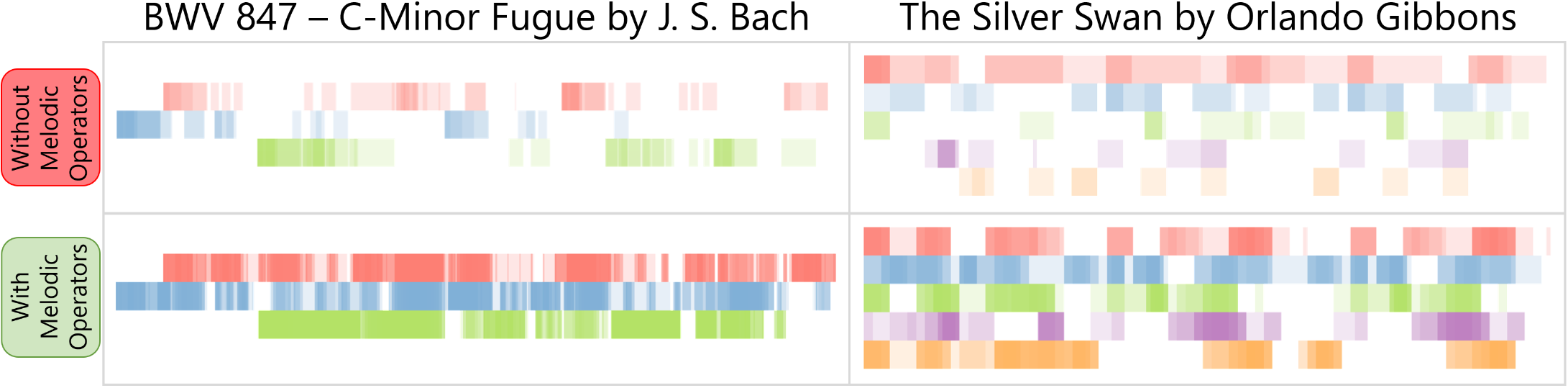} }    
    \vspace*{-12pt}
    \caption{ 
    Direct comparison of heat maps based on selected melodic patterns from all participants without the operators (top, red) and with the operators (bottom, green) for \smA~(left) and \smB~(right).    
    }
    \vspace*{-8pt}
    \label{fig:evaluation:fugue_swan_comparison}
\end{figure}

\noindent The four pattern identification heatmaps displayed in \autoref{fig:evaluation:fugue_swan_comparison} for both conditions and music sheets separately confirm that for both pieces the amount of patterns with the operator is a magnitude higher than without the melodic operators available.
Especially in the voices 3-5 of the ``Silver Swan'' the detected patterns are more clearly to see.

\subhead{NASA TLX \& SUS Questionnaire --} 
The primary purpose of the TLX questionnaire between the two conditions was to estimate whether the addition of the melodic operators would deteriorate the overall effort of the analyst since extended automation can lead to overstraining.
The average task load index for both conditions is between 60 and 75 without any salient differences, indicating that the availability of the operators leads to at least double as many patterns as without.
We present further details in the supplementary material.

\section{Discussion}

Subsequently, we discuss the implications of our system design, the major findings from our user study, the current limitations of the research, and the opportunities for future work.

\subsection{Take-Home Messages}

\subhead{Atomic Operators and Mixed-Initiative Analysis --} 
The concept of atomic operators and their combination into complex operators can be highly valuable for information visualization research. 
Our semi-automatic solution demonstrates how integrating computational techniques with human control and guidance can result in more scalable and effective pattern detection. 
This mixed-initiative approach retains human agency in the analysis process and can be adapted to other domains, such as text or bio-genome data analysis.

\subhead{Study Insights and Scalability --} Our user study showed that the use of melodic operators significantly increased the number of patterns identified with low cognitive effort. 
At the same time the familiarity with the sheet music view offered a suitable starting and reference point during the subsequent analysis.
This demonstrates the potential for scaling up complex problem-solving in information visualization while still retaining human involvement and expertise in the analysis process.

\subhead{Methodology Transfer to Other Domains --} The techniques we employed can be transferred to other data analysis contexts, such as text or temporal data. 
The combination of atomic operators, visual representations, and interactions can provide a solid foundation for tackling intricate pattern detection tasks across domains, fostering interdisciplinary collaboration and innovation in information visualization research.

\subhead{Collaborative Research --} Embracing collaborative research between different fields, such as musicology and visualization, can lead to the development of innovative techniques that support operator chaining and the simultaneous analysis of multiple musical characteristics. 
This allows for a deeper understanding of complex musical structures and relationships, ultimately fostering interdisciplinary knowledge exchange and providing novel insights in the field of visual musicology.

\subsection{Limitations and Potential Extensions}

Our results are limited by the amount of participants, the shortness of their interactions, and the broad spectrum of participants. However, the current insights already show that our system matches the initial requirements while indicating the following needs for future extension.
The addition of audio playback could enhance the user experience and understanding of the analyzed music.
The scalability of \toolname could be improved by increasing the performance for applying operators. 
The currently used colors, which rely on transparency and blended colors, could make it difficult to interpret the relationships between applied operators. 
Currently, our prototype supports only a preselected number of pieces, and an option to import custom sheet music could improve its applicability.
In the MTG, individual patterns resulting from the operators are currently combined into single nodes which could be extended. 
Adding parameterization for the operators may provide greater flexibility supporting additional use cases.
Yet, the length of a proposed pattern is fixed with no operator available to search for meaningful sub-sequence patterns.
As of now, user interaction is mandatory in the current implementation, as automatic pattern suggestions are not yet available. 
Future versions of \toolname could learn from user interactions and suggest patterns for improved analysis based on previous selections.

\subsection{Open Research Opportunities} 
The interdisciplinary research between musicology and visualization and data analysis at the intersection of visual musicology~\cite{framingvisualmusicology2019} provides compelling opportunities to uncover new insights into musical compositions, enabling scholars to explore and analyze complex melodic patterns through the fusion of traditional musicological methods and cutting-edge visual and data-driven techniques.

\subhead{Melody--Lyrics Relationships and Genre Comparison --}
Exploring the interconnections between lyrics and melodic patterns can reveal deeper meanings and artistic intentions behind compositions. 
Additionally, comparing various genres to identify unique and common melodic patterns can provide insights into how composers use complex melodic combinations in different forms of music, thus shedding light on the stylistic characteristics and conventions of each genre.

\subhead{Temporal Evolution of Melodic Motifs --}
Using abstract visualization to demonstrate the temporal evolution of melodic motifs can help researchers understand how specific melodic patterns have been employed by different composers across multiple epochs and cultures. 
To achieve this, it involves developing visualization techniques that track the evolution and adaptation of melodic motifs over time.

\subhead{Advanced Visualization Techniques --}
Developing novel visualization techniques for abstract melodic analysis could help identify differences between motifs and themes across composition collections. 
This would enable researchers to easily navigate between abstract representations and specific melodic patterns~\cite{janicke_close_2015} while facilitating a better understanding of the structure and development of various melodies.

\subhead{Leveraging Intelligent Pattern Recognition --}
Utilizing state-of-the-art machine learning and AI-based techniques bears significant potential for automated pattern recognition and classification in melodic analysis. 
Leveraging advanced algorithms and methods could improve the accuracy and efficiency of identifying melodic patterns in single pieces and whole composition collections.

\section{Conclusion}

We presented \toolname, an interactive application for semi-automatic melodic pattern investigation in digital sheet music. 
Integrating musicology experts, we developed an interface with connected views like the Melody Operator Graph and Voicing Timeline for interactive pattern exploration. 
\toolname uses eight atomic operators for effective melody analysis.
Our user study with 25 participants showed the tool's usefulness and effectiveness, doubling pattern identification compared to manual analysis. Users experienced improved pattern identification and interpretation. 
\toolname overcomes fully automated approach limitations by involving analysts for transparent and accurate melodic pattern investigation, advancing interdisciplinary collaboration between information visualization and musicology.

\clearpage

\bibliographystyle{abbrv-doi-hyperref}

\bibliography{_bib}

\end{document}